\documentclass[journal]{IEEEtran}
\usepackage{fancyhdr}
\usepackage{amsmath,amssymb}
\usepackage{amsfonts}
\usepackage[dvips]{graphicx}
\usepackage{dsfont}
\usepackage{balance}

\usepackage[hidelinks]{hyperref}    

\usepackage{color}
\usepackage{pgfplots}           
\pgfplotsset{compat=1.16}
\usepackage{enumerate}
\usepackage{graphics}
\usepackage{subfigure}
\usepackage{cite}
\usepackage{mathrsfs}
\usepackage{stfloats}
\usepackage{xfrac}

\usepackage{hyperref}
\hypersetup{
    colorlinks=true,
    linkcolor=blue,
    filecolor=magenta,      
    urlcolor=gray,
    pdftitle={Overleaf Example},
    pdfpagemode=FullScreen,
    }

\usepackage{tikz}
\usepackage{mathdots}
\usepackage{yhmath}
\usepackage{cancel}
\usepackage{color}
\usepackage{siunitx}
\usepackage{array}
\usepackage{multirow}
\usepackage{textcomp}
\usepackage{gensymb}
\usepackage{tabularx}
\usepackage{booktabs}
\usepackage{graphicx}
\usepackage{xcolor}

\usepackage{makecell}

\begin{document}
\title{Multiple Access in the Era of Distributed Computing and Edge Intelligence}

\author{Nikos G. Evgenidis, Nikos A. Mitsiou, \IEEEmembership{Graduate Student Member, IEEE}, Vasiliki I. Koutsioumpa, \\ Sotiris A. Tegos, \IEEEmembership{Senior Member, IEEE}, Panagiotis D. Diamantoulakis,~\IEEEmembership{Senior Member, IEEE}, \\ and~George K. Karagiannidis,~\IEEEmembership{Fellow, IEEE}
\thanks{The authors are with the Department of Electrical and Computer Engineering, Aristotle University of Thessaloniki, 54124 Thessaloniki, Greece  (e-mails: nevgenid@ece.auth.gr, nmitsiou@auth.gr, vasioakou@ece.auth.gr, tegosoti@auth.gr, padiaman@auth.gr, geokarag@auth.gr).}
\thanks{G. K. Karagiannidis is also with the Artificial Intelligence \& Cyber Systems Research Center, Lebanese American University (LAU).}
}

\maketitle

\begin{abstract}
This paper focuses on the latest research and innovations in fundamental next-generation multiple access (NGMA) techniques and the coexistence with other key technologies for the sixth generation (6G) of wireless networks. In more detail, we first examine multi-access edge computing (MEC), which is critical to meeting the growing demand for data processing and computational capacity at the edge of the network, as well as network slicing. We then explore over-the-air (OTA) computing, which is considered to be an approach that provides fast and efficient computation of various functions. We also explore semantic communications, identified as an effective way to improve communication systems by focusing on the exchange of meaningful information, thus minimizing unnecessary data and increasing efficiency. The interrelationship between machine learning (ML) and multiple access technologies is also reviewed, with an emphasis on federated learning, federated distillation, split learning, reinforcement learning, and the development of ML-based multiple access protocols. Finally, the concept of digital twinning and its role in network management is discussed, highlighting how virtual replication of physical networks can lead to improvements in network efficiency and reliability.
\end{abstract}    

\begin{IEEEkeywords}
Next-generation multiple access, multi-access edge computing, over-the-air computing, semantic communications, machine learning, digital twinning
\end{IEEEkeywords}


\section{Introduction}
\IEEEPARstart{T}{he} next generation of wireless communication networks is expected to provide intelligence, limitless connectivity and full synchronization of the physical and digital worlds, paving the way for the development of novel applications both indoors and outdoors. Specifically, the key performance indicators (KPIs) expectations of academia and industry are converging toward an ever-increasing deployment of machine-type nodes in 2D/3D service areas (10 Gbps/m$^3$), and the need for even higher reliability ($10^{-9}$ frame error rate), lower latency (0.1 msec), higher data rates (1 Tbps), higher energy efficiency (1 pJ/bit). 

However, except achieving higher data rates, lower latency, and more connected devices, future wireless networks can be seen as a distributed computing platform and, thus, purely communication metrics are not sufficient to describe their performance. In more detail, using wireless networks as a flexible and scalable computing platform radically changes the design and optimization of multiple access schemes, since the aim is to efficiently process the users' data, either independently at any of the available processing units of by performing calculations using as input the datasets of multiple users.  In addition, due to the complex nature of future wireless networks, they are expected to integrate native artificial intelligence (AI), which will facilitate the optimal use of communication and distributed computing resources, as well as the provision of services that are based on distributed artificial intelligence. 




\subsection{Background on multiple access}

To meet the stringent requirements of next-generation wireless networks, a key challenge is the development of advanced multiple access schemes, known as next-generation multiple access (NGMA). NGMA aims to efficiently and intelligently connect numerous users and devices using available wireless resources. Historically, multiple access schemes have played a critical role in wireless communications. These schemes are broadly divided into two types: orthogonal and non-orthogonal transmission strategies \cite{Liu2022}.

Orthogonal strategies, which include frequency division multiple access (FDMA) in 1G, time division multiple access (TDMA) in 2G, code division multiple access (CDMA) in 3G, and orthogonal frequency division multiple access (OFDMA) in 4G, allocate different frequency, time, or code resources to each user. These schemes are simple and reduce interference, making them popular in practical systems. However, as the number of wireless devices increases and spectrum availability remains limited, these strategies become less efficient. They have low spectral efficiency and can only support a limited number of users due to rigid resource allocation.

In contrast, non-orthogonal transmission strategies, which have attracted significant research interest, allow multiple users to share the same resource blocks. This approach requires additional techniques at the transmitter and receiver level to manage the resulting interference. These techniques include superposition coding, rate splitting \cite{Tegos2022, Clerckx2020, Xiao2023}, successive interference cancellation (SIC), and message passing. Despite increasing complexity at the transmitter and receiver, non-orthogonal strategies offer significant advantages. They support massive connectivity, achieve high spectral and energy efficiency, and ensure user fairness \cite{Tegos2023}.

NGMA can also be categorized into multi-tool NGMA, multi-concept NGMA, and multi-functional NGMA, each of which focuses on different aspects of wireless communication \cite{chen2023signal}.
Multi-tool NGMA emphasizes the dynamic allocation of available resource blocks using either orthogonal multiple access (OMA) or non-orthogonal multiple access (NOMA). This approach focuses on the efficient use of different types of resource blocks to optimize network performance, with OMA and NOMA offering different resource allocation strategies.
On the other hand, multi-concept NGMA aims to enhance network capabilities by integrating NGMA with other advanced transmission concepts, including but not limited to technologies such as cell-free massive multiple-input multiple-output (MIMO), reconfigurable intelligent surfaces (RISs), full-duplex relaying,  millimeter-wave (mmWave) communications, and heterogeneous networks. By combining NGMA with these cutting-edge technologies, multi-concept NGMA can create synergistic effects that lead to significant improvements in network performance and the ability to serve a large number of users simultaneously.
In addition, multi-functional NGMA focuses on designing wireless signals to support multiple functions. This involves the joint design of signal waveforms and signal processing techniques. For example, this approach is particularly relevant in applications such as integrated sensing and communication (ISAC). In ISAC, the superposition of communication and sensing signals represents a form of non-orthogonal resource sharing that is consistent with the principles of NOMA. This multi-functional approach in NGMA allows for the simultaneous handling of different wireless communication tasks, such as sensing, localization, and data transmission, within the same spectral resources.

Multiple access in wireless networks can also be divided into contention-free (CF) and contention-based (CB) categories, each with different resource allocation methods. CF multiple access uses a coordinated approach where network resources are allocated in a planned and orderly manner. This method is particularly favored in enhanced mobile broadband (eMBB) applications where high spectral efficiency is the primary goal. The structured nature of CF allows optimal use of the spectrum, making it suitable for high data rate applications such as streaming video and large file transfers in eMBB scenarios.
On the other hand, CB multiple access uses an opportunistic resource allocation strategy. This approach is more flexible because it allows devices to access network resources as they become available, creating a kind of competition among devices for resources. This method is particularly advantageous in massive machine-type communications (mMTC), a domain characterized by a large number of devices, such as in internet of things (IoT) networks. In mMTC, the sporadic and unpredictable activity patterns of devices make centralized coordination less practical. The opportunistic nature of CB helps to efficiently manage the sporadic data transmissions typical of IoT environments, where devices may only need to transmit small amounts of data infrequently.
Considering these aspects, OMA schemes are predominantly used in current eMBB applications due to their efficiency in handling high data rates and their ability to maximize spectral utilization. In contrast, for mMTC applications, where the key challenge is managing a large number of intermittently active devices, random access schemes are emerging as the preferred solution. Random access  approaches provide the flexibility and scalability needed to meet the unique requirements of mMTC environments, such as those found in large-scale IoT networks. These schemes allow a large number of devices to access the network in an uncoordinated manner, which is ideal for handling the irregular traffic patterns typical of mMTC scenarios \cite{mMTC,Tegos2020,Choi2017,Choi2018,tyrovolas2023}.

In the context of random access, grant-based random access and grant-free random access are two approaches to wireless communications, each with its own set of characteristics and applications. In grant-based random access, the base station (BS) plays a central role in scheduling and authorizing users through a four-message handshake process. This method ensures reliable data transmission by maintaining synchronization and proper resource allocation. The structured approach of grant-based random access is advantageous in scenarios where reliable transmission is critical, despite the associated higher signaling overhead and potential for increased latency in high-traffic environments.
Grant-free random access provides a more streamlined process that uses a two-step random access procedure. In this approach, active users transmit preambles and data directly to the BS without waiting for prior authorization. This reduces latency and control-signaling overhead, making grant-free random access suitable for scenarios where speed and efficiency are priorities. While this can lead to increased signal processing complexity at the receiver, grant-free random access provides a balance between efficiency and simplicity in many use cases.
Unsourced random access is another approach designed to further streamline the process for scenarios requiring massive connectivity \cite{Polyanskiy2017, Che2023}. In unsourced random access, all potential users share a common codebook, and the BS decodes a list of messages rather than identifying individual active users. This eliminates the need for preambles, reducing overhead and minimizing collision frequency. Unsourced random access is particularly advantageous in NGMA scenarios such as mMTC, where the focus is on accommodating a large number of users with sporadic and short traffic. In this context, fast uplink grant has also been proposed and used, which requires a one-way handshake and allows only devices that have received an uplink grant to transmit data, resulting in reduced access delay and collisions \cite{Ali2020, Mitsiou2023}.

The field of NGMA is in its early stages, and there is still a lot of room for research in a number of important areas. One key area is the development of new multiple access schemes. Techniques such as NOMA and space division multiple access (SDMA) are being explored. These new schemes aim to improve on traditional methods by offering higher bandwidth efficiency and better connectivity, which are critical for the evolving demands of wireless networks. Another important area of research is the development of new techniques that can enhance NGMA systems. This includes exploring the potential of RISs, which can adaptively modify electromagnetic waves for better signal transmission. In addition, advances in random access protocols and more sophisticated modulation and channel coding schemes are being investigated to improve network performance and reliability. Finally, advanced machine learning (ML) tools and big data analytics are becoming more prominent in NGMA research, promising to address complex challenges in NGMA by enabling more intelligent and adaptive network solutions. Through the use of ML and big data, researchers can develop systems that are better equipped to handle the diverse and dynamic nature of future wireless communication needs. Overall, the combination of new multiple access schemes, innovative techniques, and the application of ML and big data is essential for the progress and success of NGMA.


\begin{table*}[t]
    \caption{NGMA in the context of integrated communication and computing: Objectives and enablers.}
    \label{definitions}
    \centering
    \resizebox{\textwidth}{!}{
    \begin{tabular}{c|c|c}
        \hline \hline
        \textbf{Objective}
        & \textbf{Integrated resource communication efficiency (IRCE)}
        & \textbf{Integrated resource distributed computing efficiency (IRDCE)}
        \\ \hline \hline
        \textbf{Definition} 
        & \makecell{Level of achieving a communication objective per number of \\ communication and computing resources}
        & \makecell{Level of achieving a computing objective per number of \\ communication and computing resources} \\ \hline
        \multirow{7}{*}{\textbf{Enablers}}
        & ML-aided multiple access protocol design
        & Higher IRCE \\ 
        & Waveform design
        & Over-the-air computing \\
        & Semantic communications
        & Multi-access edge computing
        \\ 
        & Proactive resource allocation
        & Federated learning, federated distillation, and split learning
        \\ \cline{2-3}
        & \multicolumn{2}{c}{Network slicing} \\
        & \multicolumn{2}{c}{Machine learning algorithms (e.g., deep reinforcement learning)} \\
        & \multicolumn{2}{c}{Digital twinning} \\ \hline \hline
    \end{tabular}
    }
\end{table*} 

\subsection{Contribution}
In this paper, we provide a comprehensive overview of NGMA in the context of communication and computation. In this direction, as it is briefly shown in Table \ref{definitions}, we identify the main enablers for achieving two different objectives, hereinafter termed as integrated resource communication efficiency (IRCE) and integrated resource distributed computing efficiency (IRDCE). Although both IRCE and IRDCE consider the efficient use of communication and computing resources, the respective key performance indicators are substantially different. More specifically, IRCE focuses on achieving the required communication quality of service, which can be expressed as a function of spectral/energy efficiency, latency, reliability and connectivity. On the other hand, IRDCE focuses on achieving a computing objective, e.g., maximizing number of bits of bits that are processed either locally or at the edge or minimizing the mean square error of a certain calculation on the sources' data.

We first provide an overview of multi-access edge computing (MEC), which is essential to address the increasing demand for data and computing power at the edge of the network, while also exploring network slicing. We then discuss over-the-air (OTA) computing as a scheme that shows promise for calculating a range of functions quickly and efficiently, proving particularly beneficial for real-time computing needs. Next, we analyze semantic communication as a promising approach to make communication systems more efficient by focusing on transmitting meaningful data, reducing redundancy, and thus increasing communication efficiency. The mutual relationship between ML and multiple access is also discussed, emphasizing in federated learning (FL), federated distillation (FD), split learning (SL), reinforcement learning (RL), as well as ML-aided multiple access protocol design.
Finally, we analyze digital twinning and its application to network management, providing insight into how virtual modeling of physical networks can improve network efficiency and reliability. 

\subsection{Structure}
The rest of the paper is organized as follows. In Section \ref{s:MEC}, MEC in NGMA is discussed, while Section \ref{s:OTA} analyzes the concept of OTA. Section \ref{s:semcom} discusses semantic communication and joint computing and communication for distributed ML applications is analyzed in Section \ref{s:joint}. Next, in Section \ref{s:DT}, digital twin-aided multiple access is examined. Finally, concluding remarks are provided in Section \ref{s:concl}.

\section{Multi-access edge computing} \label{s:MEC}
Next-generation wireless communications and the vision of IoT have led to a significant shift in mobile computing. The conventional centralized mobile cloud computing model cannot satisfy the strict delay requirements of emerging mobile applications due to the long distances to the end users and the backhaul bandwidth limitation, which, together with the limited battery lifetime as well as computational and storage resources of mobile end devices, have recently paved the way for the integration of the MEC, also known as mobile edge computing \cite{8016573, mehrabi2019device, liu2020toward}. The primary feature of MEC is to bring mobile computing, network control and storage in close proximity to wireless end devices \cite{8807194, 8755828}. The MEC servers can be small-scale data centers co-located with BSs, wireless access points or cellular backhaul units \cite{8723481}. Bringing intelligence closer to the end users offers several advantages. In addition to providing high-bandwidth, low-latency communications, proximity, location and mobility awareness, and real-time visibility into radio network information, MEC can enable a variety of IoT and delay-sensitive applications and services. It can provide real-time information on user location and behavior to improve quality of service (QoS) and user quality of experience (QoE), while reducing backhaul traffic and supporting wireless power transfer \cite{9817408} to mobile devices.

The fundamental feature of MEC that allows the realization of the aforementioned applications is computational offloading, i.e. the distribution of computational tasks from the computationally limited end devices to the better performing MEC servers. This attribute of MEC can offer an efficient way to prolong battery life by limiting power consumption caused by extensive computational processing, while enabling end devices to successfully support various applications, even if they would be extremely challenging to perform at their end from a computational point of view. Although the end devices may have insufficient resources for the timely execution of a computationally intensive task on their own, offloading can accelerate the required computation time by taking advantage of highly computational performing servers to perform the most intensive parts of computing. Depending on the computational load of a task, offloading can be divided in two categories:
\begin{itemize}
\item Binary offloading: A computationally intensive or relatively simple task cannot be partitioned and is offloaded to the MEC server or processed locally on the mobile device.
\item Partial offloading: A computational task is partitioned into two parts, one for local execution and one for MEC offloading.
\end{itemize}

A mobile end user's task can be represented by a tuple $\mathcal{A} = \{L,\; X,\; T_{\max}\}$, where $L$ is the task input data size in bits, $X$ is the computational workload/intensity in CPU cycles per bit, and $T_{\max}$ is the required completion time in seconds. Let us consider $\theta \in [0, 1]$ be the task offloading factor, e.g., the fraction of offloaded data to the total data, and $1-\theta$ is the amount of the locally processed task. Thus, the local computation latency can be derived as
\begin{equation} \label{d_com}
    T_{\mathrm{c}} = \frac{(1-\theta) L X}{f} ,
\end{equation}
where $f$ denotes the frequency of the CPU cycles. Furthermore, the time required to transmit the offloaded amount of task to the MEC server is given by
\begin{equation} \label{d_t}
    T_{\mathrm{t}} = \frac{\theta L}{R} ,
\end{equation}
where $R$ is the transmission rate and can be calculated according to the multiple access protocol being used. Although MEC servers have large computational capabilities compared to their computational load, their computational power is not unlimited, thus it is necessary to account for the MEC server execution time in the general design of MEC systems. To this end, the offloaded task execution time in the MEC server can be written as
\begin{equation} 
    T_{\mathrm{s}} = \frac{\theta L X}{f_\mathrm{s}} ,
\end{equation}
where $f_s$ is the CPU-cycle frequency of the MEC server. Therefore, the total latency of task offloading is the sum of the transmission latency and MEC execution latency given by
\begin{equation} \label{d_of}
    T_{\mathrm{of}} = T_{\mathrm{t}} + T_{\mathrm{s}} = \frac{\theta L}{R} + \frac{\theta L X}{f_\mathrm{s}} .
\end{equation}
Taking into account that the transmission between a user and the MEC server can occur in parallel with the local computation, an end user can perform offloading and task processing simultaneously. Therefore, the total latency in the case of partial offloading is computed as $T_\mathrm{total} = \max\{T_\mathrm{c},\, T_\mathrm{of}\}$. It should be noted that the case of binary offloading is denoted by an offloading factor equal to 0 or 1. Specifically, $\theta = 1$ corresponds to full offloading with a total latency equal to \eqref{d_of}, while $\theta = 0$ represents the case where no offloading is performed, e.g. the task is executed locally, and the total latency is given by \eqref{d_com}. In any case, the total system latency should not exceed the threshold $T_{\max}$, thus $T_\mathrm{total} \leq T_{\max}$.

As previously mentioned, mobile devices are energy constrained, so the energy consumption for local computation and offloading to MEC is another critical measure of mobile computing efficiency. The energy consumption of local task computation can be expressed as
\begin{equation}
    E_{\mathrm{c}} = \kappa_\mathrm{c} (1-\theta) L X f^2 ,
\end{equation}
where $\kappa_\mathrm{c}$ is a constant related to the hardware architecture. In the case of offloading, the energy consumption for transmitting a task to the edge server is given by
\begin{equation} \label{e_t}
    E_{\mathrm{t}} = p T_{\mathrm{t}} = p \frac{\theta L}{R} ,
\end{equation}
where $p$ is the transmission power. The energy consumption for the MEC execution can be written as
\begin{equation}
    E_{\mathrm{s}} = \kappa_\mathrm{s} \theta L X f_\mathrm{s}^2 ,
\end{equation}
but is usually omitted, as we are more interested in the energy consumption of end devices.

In addition, the time and energy required by the MEC server to send the computed tasks to the edge users are not considered, since the size of the tasks is typically very small and the capabilities of the server are large. 

\subsection{Cross-layer design in MEC} 
In the context of MEC, edge computing refers to the delivery of computing resources to end users in close proximity to and at the edge of the radio access network (RAN). Consequently, MEC demonstrates the ability to decentralize computing resources from a centralized cloud to the network edge to serve a large number of users. However, scalability concerns and server congestion may arise, especially as the workload intensity of end users gradually escalates \cite{bebortta2021adaptive, guan2019mec}. Additionally, in practical scenarios, the offloaded tasks cannot be immediately executed on the MEC server due to its limited computational capabilities \cite{meng2019closed, zhou2021distributed}. The computational tasks generated by mobile end users create a data queue that encompasses dynamic and stochastic processes such as traffic arrivals, packet service, and waiting in each user's or server's buffer. The queuing dynamics of computational tasks can have a significant impact on user offloading decisions and the overall performance of MEC. To make effective offloading decisions in MEC, end users need to understand the queuing dynamics of their upper-layer tasks in addition to their communication and computation dynamics to facilitate cross-layer optimization. 
The most typical procedure followed is the application of a threshold-based task offloading policy, while the data packets' arrival time is generally modelled by a Poisson process to capture the dynamic of data bursts in networks.

A MEC system consisting of a MEC server and a mobile device with energy harvesting components is presented in \cite{zhang2018energy}. Considering the adequacy of energy resources, the computational tasks can be executed by the local CPU to reduce the transmission delay, or by the MEC server to reduce the energy consumption, or can be dropped. Both local and remote execution delay include queuing delay. The wireless channel status affects the offloading, and an appropriate wireless channel environment must be selected to save power consumption. Offloading tasks to the MEC server can reduce the intensive computation workload of the end device while increasing the transmission delay. Therefore, a trade-off between energy consumption and execution delay should be considered. 

Another offloading decision based on users' channel gains is described in a multi-user and single-server MEC scenario where multiple users share the limited spectrum and interfere with each other \cite{zhou2021distributed}. Good communication conditions, i.e., a user's channel gain exceeds a threshold, indicate task offloading to the network edge for remote execution, otherwise the task is pushed into a queue buffer. A user's offloading decision is influenced by the others', and its offloading decision threshold can be adjusted in order to optimize the expected successful offloading rate.  A higher threshold can reduce the probability of task offloading and thus lose the benefits of edge computing, while a lower threshold can lead to a higher probability of task offloading, which increases the intensity of physical-layer channel contention and wireless interference. 

In a MEC network consisting of a finite number of active MEC servers, as demonstrated in \cite{bebortta2021adaptive}, the continuous increase in the number of incoming data packets can exceed the servers' service capacity, leading to overloaded MEC servers and a system out of steady state. A threshold value is used to indicate server capacity overflow. Since no waiting queues are formed, server bursts occur and can lead to service outages for some complex delay-sensitive tasks due to overloading of MEC resources. To overcome this issue, the ability to dynamically offload the workload to cloud computing servers is provided, and minimizing the waiting time before offloading to the cloud services is of great importance.

Since workload arrivals are often uneven among the MEC-equipped small-scale BSs (SBSs), computation offloading between peer SBSs can be enabled to harness underutilized, otherwise wasted computational resources to improve the overall system efficiency \cite{chen2018computation}. Peer offloading relies on the cooperative behavior of SBSs in sharing their computational resources as well as their energy costs, but can also cause additional delay due to network congestion.

A MEC system with a limited computation buffer at the edge server has been studied in \cite{cao2020delay}. Saturation of the computation buffer causes the task transmission to be paused until additional capacity is available in the computation buffer, while saturation of the transmission buffer results in local computation. However, with a high task arrival rate, the latency requirements of all tasks that cannot enter the transmit buffer may not be satisfied, and more communication and computation resources must be allocated. The total delay of a task consists of the waiting time for transmission, the transmission time, the waiting time for computation, and the computation time. The transmit process is linked to the state of the computation buffer. Therefore, the transmission and computation processes are coupled when analyzing the random delay characteristics.

\subsection{Multiple access protocols}
In a network with multiple end users or multiple MEC servers, data transmission to a MEC server for remote execution can be performed under several access schemes. The transmission rate plays a crucial role in the task offloading latency and energy consumption, as described by \eqref{d_t} and \eqref{e_t}, and depends on the utilized multiple access protocol. Therefore, in addition to the efficient utilization of the computation and communication resources, the appropriate selection of the multiple access scheme is crucial for the efficient use of MEC. The multiple access protocols can be used for both uplink and downlink transmissions. An uplink MEC transmission is referred to as a network with multiple end user wanting to offload their tasks to a single MEC server, while the downlink MEC transmission is considered as a user with multiple tasks being offloaded to multiple MEC servers simultaneously.

\subsubsection{OMA}
In OMA, e.g., TDMA, FDMA, users belonging to the same cell are allocated to different resource blocks and do not interfere with each other \cite{diamantoulakis2021optimal}. For the TDMA system, time is divided into slots and the duration of each time slot is selected to meet the latency requirements of the users. For the OFDMA system, the total bandwidth is divided into multiple orthogonal subchannels, and each subchannel can be assigned to at most one user.

A resource allocation strategy is proposed in \cite{7762913} for a multi-user single-server MEC network. The offloading transmission is based on TDMA and OFDMA to achieve high energy efficiency and low latency of task offloading. In the multi-server MEC network topologies presented in \cite{8533343, wang2021multiobjective}, multiple optimization objectives are accounted simultaneously, e.g., minimizing the total time and energy overhead of the MEC system and the user subscription cost. The power and computing resource allocation of the users is achieved, as well as the task offloading decision is derived.
The above works are under the assumption of perfect CSI. However, in practical wireless communication networks, obtaining accurate CSI is challenging, mainly due to the feedback delay or error in channel estimation. A multi-user MEC-enabled network for IoT and Internet of Vehicles (IoV) under imperfect CSI was studied in \cite{9187942, WU2022101867}, respectively, where different frequency blocks are assigned to end users. The proposed algorithms provide advantages in terms of weighted sum of energy consumption, satisfying delay requirements, and reducing system overhead.

\subsubsection{NOMA}
NOMA is considered a key enabler for current and next-generation networks. The key idea of NOMA is the use of superposition coding and interference cancellation techniques. Compared to conventional OMA, NOMA can allow multiple users to share the same time or frequency resources and has the potential to achieve higher spectral efficiency, lower latency, massive connectivity, and relaxed channel feedback \cite{pham2020survey, 8010756, 7842433}. These advantages are mainly due to the fact that in NOMA, users with bad channel conditions do not exclusively occupy the resource blocks. The combination of MEC and NOMA can improve user satisfaction and network performance by leveraging NOMA's spectral efficiency and latency reduction benefits with MEC's end-user, operator, and overall network efficiency benefits. In addition, the flexible scheduling, grant-free access and increasing number of users provided by NOMA, as well as the empowerment of services running at the edge, can lead to low-latency and energy-efficient communications \cite{pham2020survey}, distributed computation and reinforcement of services and applications supported by next-generation networks. It is worth noting that NOMA and MEC can be dynamically integrated with other wireless technologies, such as MIMO, massive MIMO, mmWave communications, etc., to further increase connectivity, spectral efficiency, energy efficiency, and computing capacity. 

According to \cite{ding2018impact}, NOMA-based MEC is superior to its OMA counterpart when server grouping with strong channel conditions is applied in downlink transmission, and when user grouping with diverse channel conditions is applied in uplink transmission. Under the assumption of imperfect CSI, a multi-user multi-BS MEC network was investigated in \cite{9353556} and an energy-efficient resource allocation when users utilize the NOMA protocol was performed, while in \cite{jiang2020performance} the offloading outage probability, energy consumption and throughput were investigated for a NOMA-based multi-user and single-server MEC network. The proposed schemes outperform the orthogonal counterparts. However, imperfect CSI degrades the system performance compared to perfect CSI.

\subsubsection{Hybrid NOMA}
Integrating NOMA into MEC promises to minimize energy consumption and mitigate significant delays. However, studies have demonstrated that employing NOMA alone may not necessarily enhance performance of MEC computing systems, particularly when users have different latency requirements. Therefore, a hybrid NOMA-OMA scheme has been proposed \cite{8492422, 8673584, ding2022hybrid, diamantoulakis2021optimal}. 
In this hybrid NOMA scheme, both users can transmit their messages to the MEC server utilizing the NOMA protocol in the first phase of transmission, and in the second phase, only one of the users can transmit the rest of its task using the OMA protocol. In pure NOMA, both users offload their tasks in the first phase, while in OMA, the first phase is accessed by only one user. 

Pure NOMA can outperform hybrid NOMA when there is sufficient energy for MEC offloading, while hybrid NOMA always outperforms pure NOMA in terms of minimizing energy consumption \cite{8492422}. MEC based on hybrid NOMA was shown to outperform its OMA counterpart when users have demanding latency requirements for their task offloading, while OMA-based MEC is preferred when a user's task is delay tolerant \cite{8673584}. The NOMA-based MEC scheme is not preferred for either situation. A generalized MEC offloading strategy based on hybrid NOMA was proposed in \cite{ding2022hybrid} for a MEC network with more than two users, and a multi-objective optimization problem was formulated to minimize the energy consumption of offloading. In \cite{diamantoulakis2021optimal}, it was concluded that the use of dynamic user scheduling and time-sharing during the initial transmission phase can significantly reduce latency, especially when energy consumption is kept relatively low. A partial offloading scenario with controllable CPU-clock speed was also considered, which is interesting because it involves optimizing both communication and computational resources. Increasing the CPU speed of the local processor implies a reduction in latency without significantly increasing transmission power consumption and potential interference. However, it also results in higher local power consumption. Therefore, the coordinated optimization of advanced communication protocols and computing resources holds the potential for significant improvements in terms of latency reduction and energy consumption.

\subsubsection{RSMA}
In recent years, RSMA has been investigated as a multiple access strategy for upcoming wireless communications, focusing on aspects such as non-orthogonal transmission, interference management, and rate maximization. Compared to uplink NOMA with SIC processing, the adoption of uplink RSMA offers the advantage of attaining the full capacity region of multiple access channel (MAC) \cite{485709}. However, the optimal SIC decoding order and the transmit power allocation need to be determined by exhaustive search to realize the full capacity region of uplink MAC using RSMA \cite{9257190, 8850093}. This requirement restricts the indiscriminate use of uplink RSMA in latency-critical MEC applications. Effectively applying RSMA to uplink MACs necessitates efficient interference management and low-complexity SIC processing.

Therefore, an RSMA-aided MEC scheme was introduced in \cite{10032159} to enhance the successful computation probability and reduce offloading latency within a MEC system consisting of a single MEC server and multiple randomly distributed users. Each user pair is assigned to different orthogonal time/frequency resource blocks based on distance and employs two-user uplink RSMA, eliminating interuser interference within each pair. To facilitate simultaneous offloading of paired users to the MEC server, cognitive radio (CR)-inspired rate splitting is employed, which exhibits superior performance compared to its NOMA counterpart. In \cite{9775664}, a single-server MEC network was studied, focusing on air user offload tasks using the RSMA protocol, and a joint optimization problem involving offload decisions, transmission rates, uplink power control, and decoding order was formulated. In addition, \cite{diamanti2023delay} explored an RSMA-based multi-server MEC system where users are assigned to separate frequency bands and downlink RSMA principles are applied to facilitate a single user's concurrent offloading of computation tasks to multiple MEC servers. The joint optimization of computational task assignment, radio and computational resource allocation to minimize the sum of the maximum delays experienced by users was examined.

\subsubsection{Hybrid RSMA-TDMA}
Although the use of pure RSMA in MEC systems has not been widely investigated, a hybrid RSMA-TDMA scheme for a two-user MEC network was introduced in \cite{xiao2023delay}. The offloading phase is divided into two stages. First, a CR-inspired RSMA strategy is used for users to offload their computational tasks to the MEC server. Subsequently, only a single user has the ability to offload during the second phase. The objective is to minimize the offloading delay, and the power allocation is optimized under energy and power constraints. The hybrid RSMA-TDMA scheme demonstrates superiority over its hybrid NOMA-TDMA and TDMA counterparts.

\subsubsection{Grant-free random access}
In IoT applications where devices have sporadic or bursty data to transmit and may not need to access the network constantly, grant-free random access is a suitable multiple access protocol. In traditional communication systems, devices typically have to contend for access to the network, and once granted access, they can transmit their data. However, in grant-free random access, devices can directly transmit their data without prior coordination, leading to access delay and signal overhead reduction and making them more flexible and scalable than traditional grant-based
protocols.

In \cite{9906525}, the slotted ALOHA protocol was used for the wireless powered devices (WPDs) to access the medium for binary offloading tasks to a MEC server. This protocol allows the WPDs to contend for the available slots to transmit their data to the MEC server, enabling efficient and low-complexity access for the devices. Additionally, the use of grant-free random access with adaptive rate as the multiple access protocol for offloading tasks to the edge server has been proposed in \cite{10077734}. This approach offers lower implementation costs and improved efficiency for task offloading, while the average delay and resource allocation strategy are also analyzed.

\subsection{MEC with network slicing}
In next-generation wireless communications, network slicing has emerged as a key technology that facilitates the delivery of customized services that require high flexibility and scalability. Through network slicing, a physical network can be partitioned into multiple logical networks, enabling the simultaneous support of on-demand customized services for different application scenarios on a shared physical network \cite{8016573, liu2020toward, alliance2016description}. In addition, network resources can be dynamically and efficiently allocated to logical network slices according to QoS requirements. Network slicing integrates cloud and network resources, including utilities such as storage, big data processing, RAN access, and bandwidth, collectively referred to as general MEC utilities \cite{7931566}.

Combining network slicing with MEC addresses specific requirements such as edge analytics and scalability \cite{8391395}. This integration is proven valuable for latency-critical services such as the industrial Internet, healthcare, and autonomous driving, where MEC and network slicing help reduce latency and prioritize traffic. Network slicing is also of supreme need for three generic use cases in next-generation wireless networks, including eMBB, ultra reliable and low-latency communication (URLLC), and mMTC. As mentioned in \cite{7931566}, for high-capacity mobile broadband network slices, MEC increases core network and mobile backhaul capacity through content caching at the edge and traffic offloading. In the context of automotive network slices with stringent latency and scalability requirements, the capabilities of MEC are tailored to service needs. For massive IoT services that require scalability to handle large volumes of IoT data, MEC provides storage and computation capabilities that support network slice scalability.

The effectiveness of network slicing in a NOMA-enabled MEC network was investigated in \cite{9560104}. A network slicing technique was introduced to categorize users based on their service requirements and to efficiently allocate compute and radio resources. This access scheme showed significant advantages over OMA and NOMA without network slicing in terms of latency, energy efficiency, and spectral efficiency. 



\subsection{Future directions}

Current communication protocols such as OFDM, used as the physical-layer modulation scheme in 5G mobile systems, exhibit near-capacity performance over channels with minimal Doppler effect. However, as we transition to the more dynamic and mobile environments of 5G and beyond deployments, characterized by time-varying multipath fading due to mobility, OFDM encounters significant performance challenges. Furthermore, NOMA, particularly under real-world conditions with imperfect impairments, has demonstrated degraded performance, which may not justify its use compared to OMA protocols in many use cases \cite{imperfect1, imperfect2, imperfect3, imperfect4}. Addressing these challenges and advancing beyond current limitations is crucial for meeting the KPIs of future MEC networks. An important approach in this direction is ML-based multiple access schemes design, which will be discussed in section \ref{section:mac design}. Another interesting research direction is the design of new waveforms for medium access control.  One such promising paradigm is orthogonal time frequency space (OTFS). 

Unlike traditional modulation, OTFS operates in a 2D time-frequency space, spreading signals over multiple dimensions and encoding information in the delay-Doppler domain. This representation as a 2D matrix of signal delay and Doppler shift offers advantages such as high spectral efficiency, robustness to channel variations, low latency, and accurate channel estimation. OTFS is well suited for applications such as autonomous vehicles, cellular networks, and satellite communications, since it is resilient to fading, multipath effects, and Doppler shifts \cite{assess}. However, implementing OTFS can pose challenges in terms of the required transceiver complexity, which may be beyond the capabilities of low-power devices \cite{otsm1,otsm2}.

To address these challenges, orthogonal time sequency multiplexing (OTSM) has been also proposed. Unlike OTFS, which operates in the delay-Doppler domain, OTSM multiplexes information symbols in the delay-sequency domain, where sequency represents the number of zero crossings per unit time interval. OTSM achieves this by transforming information symbols from the delay-sequency domain to the delay-time domain using the inverse Walsh-Hadamard transform (WHT) along the sequency domain. This transformation allows channel delay spread and Doppler spread to cause intersymbol interference along the delay and sequency dimensions, respectively, while remaining separable at the receiver, similar to OTFS. Notably, this separability is not achieved by OFDM, where channel delay spread and Doppler spread jointly cause interference along the frequency dimension, leading to orthogonality issues between OFDM subcarriers. OTSM stands out for its low-complexity transceiver design, which relies only on WHT, and its efficient detection, comparable to traditional baseline OFDM schemes. Additionally, OTSM demonstrates strong performance in high-mobility channels, making it suitable for very low-power wireless devices such as sensors and IoT devices \cite{otsm1,otsm2}.

Nevertheless, OTSM has received comparatively less attention than OTFS. Additionally, \cite{usc} introduced unitary-precoded single-carrier (USC) modulation as a family of waveforms that multiplex information symbols on time-domain unitary basis functions spanning the entire time and frequency plane. Notably, OTFS and OTSM, based on discrete Fourier transform and WHT, respectively, fall within the general framework of USC waveforms.  Future research on USC and OTSM can focus on further analysis of channel representation, low complexity channel estimation and detection in the delay-sequency domain. Also, reducing the peak-to-average-ratio (PAPR) of USC waveforms has yet to be thoroughly analyzed, while, adaptive modulation for USC is also crucial to support eMBB application in high-mobility scenarios. Furthermore, current OTSM research focuses mainly on single-user communication, which cannot adequately capture the KPIs of future multi-acess wireless networks. Although \cite{otsm4} took a first step towards multi-user scenarios, many open questions remain. The study of multi-antenna OTSM is essential for its integration into future networks. Beamforming techniques can enhance spatial multiplexing and mitigate interference in multi-user communication scenarios. Another promising direction is to evaluate the scalability of OTSM for mMTC, serving numerous devices with sporadic traffic patterns simultaneously. 

\section{Over-the-air computing} \label{s:OTA}
\subsection{Preliminaries}\label{sub:OTA1}
In the next-generation communication systems, new applications are expected to be utilized, highlighting the need for new multiple access schemes that are appropriate for each case. A particular example of such applications is computing based on data aggregation from multiple devices. Although the computation itself is currently done with great efficiency in fusion centers, i.e., in the processing units of the BSs, the massive connectivity of devices that is expected will pose a great challenge to real-time computing, as traditional multiple access schemes will not be able to handle the amount of incoming data. For example, TDMA will result in high latency when data aggregation is required because the fusion center (FC) can only perform computation when data is available from all devices. Similarly, FDMA techniques, while preferable in terms of latency, are inefficient in terms of bandwidth allocation. On the other hand, NOMA techniques would also be quite difficult to implement, both because of the more sophisticated hardware required and the large number of devices that would need to be connected. 

With this in mind, a multiple access scheme that has received a lot of attention lately is OTA computing.  Unlike traditional schemes, OTA computing is designed for simultaneous transmission from all devices. In addition, it is designed to reduce the amount of data processing itself, which is usually not possible with transmission alone. Therefore, OTA computing is considered an efficient wireless data aggregation technique, which makes it an attractive option for many IoT applications. As shown in the seminal works \cite{stanczak_nomo,goldenbaum_sensors, Goldenbaum_eff_comp}, OTA computing can be used to calculate the value of a wide variety of functions by utilizing the superposition property of a MAC. Assuming that there are $K$ devices in the system, the functions that can be calculated using this technique are called \textit{nomographic} and are characterized by the following decomposition rule:
\begin{equation}\label{eq:nomographic}
    f(x_1,x_2,\cdots,x_K) = \psi \left( \sum_{k=1}^{K}\phi_{k}(x_k) \right),
\end{equation}
where $f:\mathbb{R}^{K} \rightarrow \mathbb{R}$ denotes the multivariate target function of OTA computing and $\psi:\mathbb{R} \rightarrow \mathbb{R}$ and $\phi_{k}:\mathbb{R} \rightarrow \mathbb{R}, \forall k \in \{ 1, \cdots, K \}$ are suitable post- and pre-processing univariate functions, respectively. As proven in \cite{goldenbaum_robust,goldenbaum_space}, any multivariate function can be written in this form. Since many goal-driven applications require the computation of such general functions, e.g., the average temperature of an area can be calculated by superimposing the transmitted signals and a post-processing dividing factor, the use of OTA is intertwined with the computation of a function of interest for some specific goal. 

One of the distinguishing characteristics of OTA computing, unlike other traditional multiple access systems such as NOMA, is that it does not require the extraction of each user's information through sophisticated techniques such as SIC, but is based on the superposition principle resulting from the simultaneous transmission of all devices. This allows OTA computing not only to approximate any real multivariate function, but also to drastically reduce the calculation time of its target function. However, studies have proved that from an information-theoretic point of view, OTA computing's potential is maximized when analog transmission is used \cite{Gastpar_optimal} instead of the highly utilized digital transmission used by current communication systems. This means that instead of conventional performance metrics, like bit error rate, the focus is on minimizing the mean squared error (MSE) between the ideal and actual received signal at the receiver point. Although, as in analog communications, no error correction can be performed, it is important to note that digital transmission itself would suffer from a computational point of view, since the required quantization for a large number of devices will result in distortion between the ideal function to be calculated and the received one. For simplicity and without loss of generality, it is assumed here that the desired function to be calculated is the arithmetic mean of the transmitted data, i.e., the ideal and received signals are given, respectively, as
\begin{equation}\label{eq:OTAidealvsRec}
    y = a\left(\sum_{k=1}^{K} x_kb_kh_k + n\right) \hspace{1mm} \text{ and } \hspace{1mm} r = \sum_{k=1}^{K} x_k,
\end{equation}
where $h_k$ are the channels of the corresponding devices, $b_k, \ \forall k \in \{ 1, \cdots, K\}$ are the pre-processing power factors at the transmitters, $a$ is the post-processing receiver gain factor and $n$ denotes the AWGN present at the receiver with noise power $\sigma^2$. For practical reasons, the transmitter power factors are also constrained up to a maximum power $P$, i.e., $|b_k| \leq \sqrt{P}, \forall k \in \{ 1, \cdots, K \}$. Thus, assuming that the transmitted signals have zero mean and unit variance, the MSE between the received signal and the desired target function is given as 
\begin{equation}\label{eq:OTAMSE}
    \mathrm{MSE} = \mathbb{E}[|y-r|^2] = \sum_{k=1}^{K} \left| ab_kh_k-1 \right|^2 + \sigma^2a^2,
\end{equation}
where the expectation is taken with respect to the transmitted signals and noise. 

\begin{figure}
\centering
\includegraphics[width=1\columnwidth]{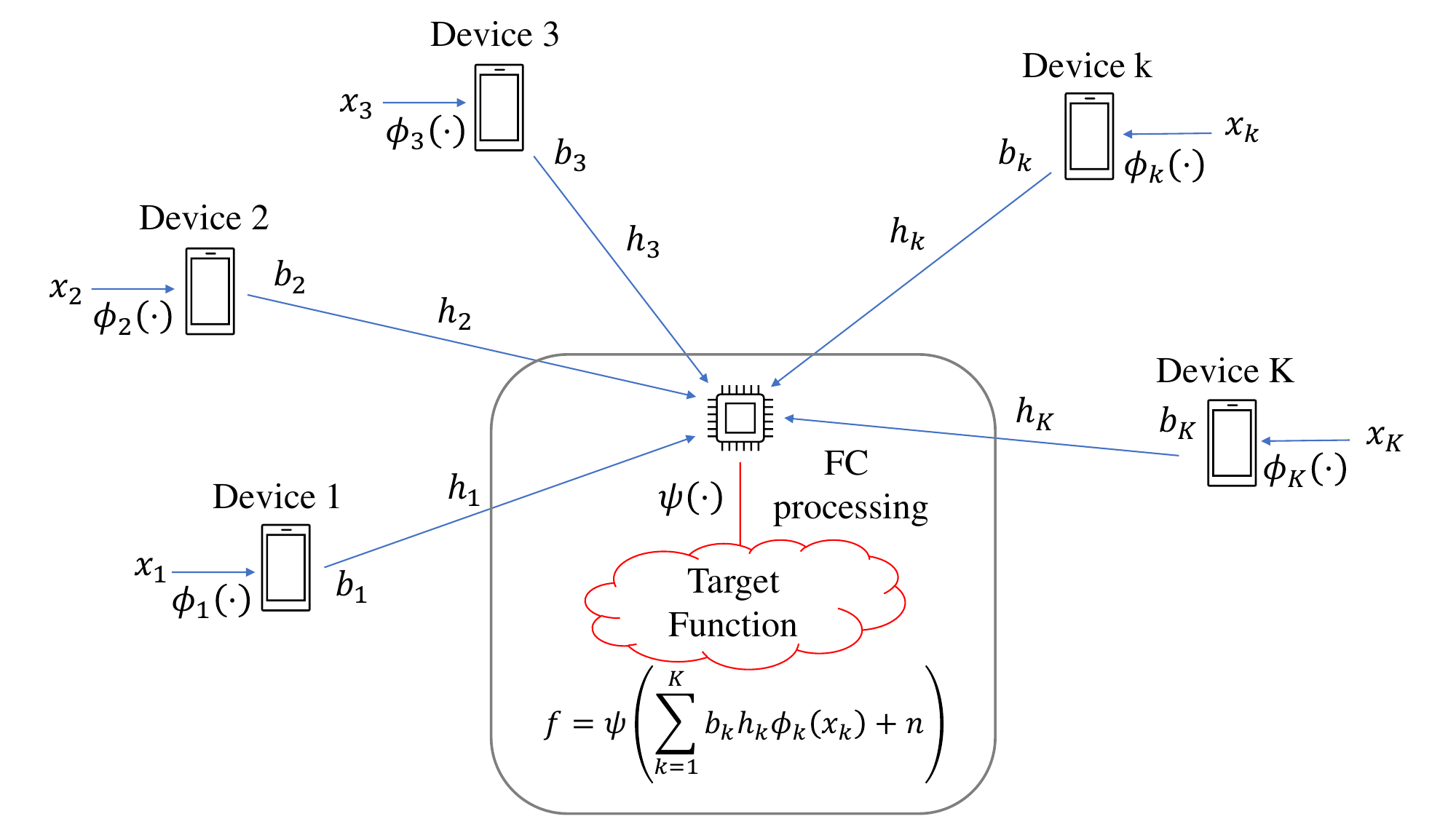}
\caption{Model of an OTA computing system consisting of $K$ transmitting devices and $1$ fusion centre.}
\label{fig:ModelSketchNew}
\end{figure}

\subsection{State-of-the-art and other technologies}\label{sub:OTA2}
To this end, many research works have emphasized on finding optimal power transmission schemes to minimize \eqref{eq:OTAMSE} mainly under perfect knowledge of the system environment, i.e., the channel conditions. In \cite{vucetic,cao_optimal}, the optimal power allocation was found under individual power constraints at each user, yielding a policy that utilizes maximum power transmission for the weakest channel devices and inverse channel transmission for the rest of them. As proven, defining the number of devices that utilize maximum power as 
\begin{equation}\label{eq:criticalNoOTA}
i^* = \underset{1 \leq i \leq K}{\mathrm{argmax}} \{a_i\},
\end{equation}
where the receiver gain factor is given as
\begin{equation} \label{eq:recfactorOTA}
a_i = \frac{\sqrt{P} \sum_{k=1}^{i} |h_k|}{\sigma^2 + P \sum_{k=1}^{i} |h_k|^2},
\end{equation}
the MSE in \eqref{eq:OTAMSE} can be rewritten as 
\begin{equation}\label{eq:opimalFullCSIOTA}
    \mathrm{MSE} = \sum_{k=1}^{i^*} \left| ab_kh_k-1 \right|^2 + \sigma^2a^2,
\end{equation}
from which the MSE performance is directly evaluated. The same problem  was investigated in \cite{OTA_sum_power} for a global power constraint over all OTA computing users. Furthermore, a great deal of analysis has been made on the assumption of MIMO users for an OTA computing system. More specifically, in \cite{mimo13}, a joint hybrid beamforming of a MIMO OTA computing system was analyzed to minimize the MSE in an effective way in terms of resources management. In \cite{mimo14}, an energy efficiency maximization was studied to minimize the consumption of power at the transmitting devices considering a minimum MSE threshold constraint. Moreover, in \cite{MIMO1}, a multiple target function scenario for parallel simultaneous computing through MIMO utilization was studied. Furthermore, in \cite{MIMO2}, a zero-forcing beamforming transmit scheme was proposed along with a selection combining implementation at the receiver side and the performance of the system was measured by an MSE outage metric. Finally, in \cite{MIMO3}, MIMO was considered as a means to enable multiple functionalities such as integrated sensing, communication, and computing through OTA computing and various algorithms that support the coexistence of these functionalities were examined. 

In addition to the fundamental optimization problems and their generalizations through MIMO, OTA computing has also been investigated in conjunction with other emerging technologies. Among others, RISs have been investigated as a means to implement OTA computing with higher performance gains \cite{Bouzinis_ris}. The main idea behind RIS integration with OTA computing is that it can provide coverage for devices that would otherwise not be able to participate in the system for computing while also improving the channel conditions of all devices through the additional multipath available for exploitation. Closely related to this is the use of unmanned aerial vehicles (UAVs) as a means of enabling OTA computing systems if the channel conditions between the devices and a ground-located FC pose severe performance degradation, for example due to extremely high fading resulting from multiple obstacles. With this in mind, in \cite{uav_enable}, a UAV trajectory optimization problem was studied to minimize the MSE and, in \cite{uav_enable_imperfect}, a similar scenario was considered under channel state information (CSI) imperfect knowledge and the UAV acting as an FC itself. 

\subsection{OTA computing and ML}\label{sub:OTA_ML}
In addition to the optimization problems and collaboration with other technologies described above, the ability of OTA computing to reduce processing time during computing has made it an attractive option for distributed learning techniques, such as FL \cite{OTA_FL_joint,cao_optimized_FL_2022}. In this scenario, OTA computing assists in updating the global parameters of deep neural networks (DNNs), where local terminals are responsible for updating the parameters of local DNNs and the FC is responsible for updating the global parameters, which are then passed back to the local terminals for another training round. OTA computing can provide a timely update at the FC, thus reducing the overall training time. It also is important to highlight the fact that although there is a gap between the desired calculation and the received signal, it has been shown that the utilization of OTA computing leads to convergence of training \cite{nam_conv}, which is the task of interest. Furthermore, it is worth noting that a digital version of over-the-air that performs FL with one-bit quantized gradients, called OBDA, was proposed in \cite{One_bit_gunduz} and its performance along with its convergence capability was studied for various scenarios, including pure AWGN channel, fading channel with perfect CSI and fading channel with imperfect CSI. Since successful training is the goal of FL, the convergence of FL through the OTA computing implementation showcased in \cite{One_bit_gunduz,nam_conv} is extremely important.

\subsection{Practical implementation and challenges}\label{sub:OTA_practical}
Perhaps the biggest drawback of OTA computing is its use of analog transmission. Although this is optimal in terms of MSE performance, modern communication systems are intertwined with digital transmission and are thus unable to support a practical implementation of OTA computing in its current form. With this in mind, it is crucial that future research focuses on finding ways to adapt OTA computing in digital systems. This opens up a new area of research for OTA computing, since even fundamental principles need to be studied, such as which type of pulses are preferable to use and how the performance is affected by interpulse interference. In this front, a ML model called ChannelComp \cite{channelcomp} that tries to find appropriate signals to be transmitted for a digital implementation of OTA computing has been proposed as a possible solution, however, further research is needed to provide insights about the performance of this DNN for a greater number of devices and different target functions.

One of the major advantages of OTA computing over other multiple access schemes is its ability to enable massive connectivity. However, because of this feature, any imperfections on a single-user basis can potentially cause large impairments due to the large number of connected devices in the OTA computing system, which will lead to incremental performance degradation. In general, these imperfections can be products of electronic impairments or natural phenomena such as the presence of noise. To address this practical problem, it is necessary to study the effect of imperfections in OTA computing and to find new optimal policies that can mitigate their effects. For example, as shown in \cite{me_OTA}, imperfect CSI knowledge requires a different optimization problem to be formulated, as the direct application of the technique proposed in \cite{vucetic} results in
\begin{equation} \label{eq:MSEimperfectCSI}
\begin{aligned}
\mathrm{MSE} =& \underbrace{\sum_{k=1}^{i^*} \left| \left(a \sqrt{P}\frac{{h_k'}^H}{|h_k'|} h_k -1 \right)\right|^2}_{\text{Full Power Terms}} + \sigma^2|a|^2  \\ &+ \underbrace{\sum_{k=i^* + 1}^K \left| \left(\frac{h_k}{h_k'} -1 \right)\right|^2}_{\text{Inverse Channel Terms}}.
\end{aligned}
\end{equation}
In \cite{me_OTA}, a thorough analysis of the performance of OTA computing under imperfect CSI knowledge was investigated and an optimal power allocation policy was proposed to improve it. As shown in \eqref{eq:MSEimperfectCSI}, these imperfections will cause performance degradation and thus need to be addressed to mitigate their effects. Similar studies on other types of imperfections, such as signal misalignment or phase noise, also need to be investigated in order to create a general OTA computing system model that can be adapted to practical operating conditions. In addition to studying the performance of these scenarios, the scope of future studies should also include the study of resource utilization. This is extremely important because from a practical point of view, certain performance improvements may be plausible, e.g., by re-estimating the channels for more accurate CSI, but at the cost of valuable resources of the OTA computing devices \cite{me_OTA}. The use of appropriate utility functions, such as the \textit{retransmission policy cost (RPC)} \cite{me_OTA}, which jointly consider MSE and resources, can be useful in this direction. RPC is defined as
\begin{equation} \label{eq:RPC}
\zeta (k) = \mathcal{C}(k)^{m_1} \left(\frac{\mathbb{E}[\mathrm{MSE}_k]}{K} \right)^{m_2},
\end{equation}
where $\mathcal{C}(k)$ denotes the resource cost required for $k$ retransmissions, $\frac{\mathbb{E}[\mathrm{MSE}_k]}{K}$ symbolizes the MSE for $k$ retransmissions, and the primary concern over available resources is considered to be the selected cost. Furthermore, the parameters $\{{m_1},\,{m_2}\}\in\mathbb{R}^2$ are considered as weights for the selected cost and the MSE, respectively. The parameters can be chosen in different ways to prioritize either the resource cost for retransmissions or the maximum error tolerance. By definition, RPC can be modified, i.e., different costs can be considered, and its weight parameters can be appropriately adjusted whenever new channel information becomes available to better capture the current state and needs of the system, making it an interesting utility function to consider in practical OTA computing designs. Similar utility functions can be studied to address MSE performance in combination with other valuable resources of the system depending on the state of the latter.  

As this discussion indicates, it is of paramount importance to generalize existing OTA computing resource allocation policies to more practical situations. To achieve this, we need to consider generalizations related to the availability of the resources themselves, apart from system models that encapsulate realistic operational parameters. For example, the most commonly studied models in the literature are based on power constraints that are identical for all devices, as described in the presented system model, or take all devices into account, i.e., $\sum_{k=1}^K |b_k|^2 \leq P_{tot}$, where $P_{tot}$ is an overall power constraint of all devices. Note, however, that in many cases such constraints may not be accurate, especially when energy efficiency is also of interest. In such cases, it may be more realistic to allocate resources according to their current availability, e.g., a battery-powered device may need to adjust its transmit power levels to extend its lifetime. These factors also need to be taken into account to provide more accurate performance simulations of OTA computing systems for operational use. 

\subsection{Future directions}\label{sub:OTA3}
As explained so far, OTA computing is a scheme that can be used in MAC, but until now we have assumed that the number of devices participating in such a system is known and fixed. However, in many scenarios, a device may want to communicate or join a system randomly, which is common in practice to achieve greater energy efficiency by not wasting resources unnecessarily. For example, we can consider a scenario where a device is inactive and thus does not transmit data, but when it is active it needs to transmit its current state. In these cases, the number of devices in the system is not fixed, but varies randomly, which would change the optimal power allocation policy of \eqref{eq:OTAMSE}. Thus, the study of OTA computing in random access environments would be of great interest, especially considering the fact that, unlike other random access channels, interference is actually useful in OTA computing since superposition is desired. 

Another interesting topic to study in OTA computing systems is security. In one-to-one communication, a single device transmits its data to a BS, so any eavesdropper can receive the signal from that device and try to decipher information about its data. To address this, different layers are used to handle information at different levels and certain protocols can act as a protection by encrypting the original data of the device. On the other hand, in OTA computing systems, data information can still be intercepted by eavesdroppers, but the case differs from the one-to-one scenario in two fundamental ways. First, the pre-processing function of each device already acts as a protection on the original data, but most importantly, the superposition of all signals from the OTA computing devices does not allow the extraction of individual data information from the devices. Even in digital form, OTA computing could prove vulnerable to eavesdroppers due to the large number of devices it can support, as extremely sophisticated hardware would be required to even attempt to process individual device data. With this in mind, OTA computing could be an interesting way to make computing not only more timely, but also more secure than other techniques. 

Because of its great potential for computing, another intriguing topic that could take on substance is multi-function computing. Since OTA computing is very resource-efficient compared to other multiple access protocols such as TDMA and FDMA, it would be interesting to combine OTA computing with one of these protocols to enable parallel computation of multiple target functions. For example, FDMA techniques could be used to compute different functions from different subcarriers. Although this concept seems easy to study, since orthogonality in time or frequency could be used, in practice orthogonality is rarely achieved. For this reason, multi-function computation would require the study of imperfections that would cause interference between the different functions to be computed. Since this type of parallel computing through OTA computing could greatly reduce the amount of traffic and free up valuable resources for other network applications, multi-function computing could be another direction for future studies.

\section{Semantic communications} \label{s:semcom}
In next-generation communication systems, one of the main goals will be to achieve extremely high data rates to handle the amount of data trafficking along the networks \cite{Ericsson2022}. The most common way to increase data rate is usually to increase bandwidth, which increases the capacity of digital communication systems. However, this comes at the cost of reducing available bandwidth resources from other applications that need to be processed by the network. In addition, to increase the bandwidth allocated to an application, it is necessary to support larger carrier frequencies, i.e., to move to higher frequencies. As a result, devices must be equipped with the hardware to support this transition. For example, larger carrier frequencies require smaller antenna sizes and more sophisticated electronics for oscillators, which can be challenging for widespread commercial use. 
While there are ways to overcome these challenges, at least when not many device changes are required, this view fails to capture the characteristics of the data being transmitted that can be exploited to improve network efficiency and performance.

Many applications in next-generation networks will be more interested in the correct inference for a task based on the received data than in the error-free, accurate reception of the data itself. Taking advantage of the rise of AI, it is possible to consider inference as the goal of communication by implementing DNNs that are trained to provide appropriate representations of the original data that can be correctly interpreted by the receiver, even if they differ from the original data. Semantic communications can change the way communication is achieved by taking into account the disparity between the underlying meaning of transmitted and received messages as explained above. Shannon and Weaver, in their seminal work \cite{shannon}, distinguished three levels of communication, the first of which, level A, addresses the technical problem of how accurately information is transmitted and is the one that has driven communication to date. Taking into account the meaning of the transferred data, two more levels are identified, level B, which is the semantic problem and is related to the amount of information a message can convey, and level C, which is the effectiveness problem. By incorporating underlying semantic information into the conventional communication paradigm, semantic communication aims to improve data exchange \cite{shannon}. To take advantage of the semantics of messages, information is shared a priori among all network participants in the form of knowledge bases that contain semantic relationships of the data to be transmitted. In this way, semantic communications can also increase reliability, since an error at the bit level does not always translate into an error at the semantic level. In addition, semantic communications can reduce the amount of redundant information related to the task at hand by transmitting only the necessary information and thus not overloading the network with unneeded data. Although the concept of semantic communications is quite old, it has recently come back into the limelight due to recent advances in the field of DNNs, such as natural language processing and image processing, which allow it to take shape. These developments enable the recognition of contextual relationships within text and images. These relationships are then used to extract semantic information from new data to be transferred. Another powerful tool that has enabled the use of semantic communication is the joint source-channel coding (JSCC) scheme, which provides robustness against channel fading conditions and semantic noise. The use of JSCC schemes both reduces the amount of information to be transmitted and achieves better performance. Thus, current AI techniques provide realistic ways to implement semantics-aware systems.

Apart from the use cases enabled by DNNs themselves, the study of multiple access schemes in the context of semantic communications is of great interest. In contrast to conventional techniques, in semantic communications, multiple access must be incorporated into the training process of the semantic receivers to achieve reliable performance. While in orthogonal multiple access schemes, such as TDMA or FDMA, semantic communications could be adopted and used as they currently are in schemes such as NOMA, where users also act as interference to each other, the training must take them into account, thus differentiating the semantic transceiver from its originally proposed design. For this reason, multiple access is of great interest to integrate semantic communication in next generation systems. 

Regarding the modeling and formulation of semantic communications, some general concepts based on information theory exist \cite{towards,SemComOverview}, however most of the relevant literature focuses on their applications. The most characteristic of them are related to the extraction of information from images, speech, and text. Due to the different inference tasks that are performed as a result of the different types of data, we distinguish them as the basic application cases that semantic communications have focused on so far. Although in principle the concept behind this type of communication is the same for each case, different DNN architectures and performance metrics are required to suit the data characteristics of each category.

\subsection{Semantic communications for text inference}\label{sub:semText}   
Text data is frequently responded to in many everyday applications and is thus the focus of many works. Due to the fact that the conveyed meaning of the original data rather than the exact data itself is of primary interest, it is challenging to find appropriate performance metrics to measure the likelihood of the received and original data. For text transmission, two different metrics have emerged as the most representative for performance evaluation. One of them is the bilingual evaluation understudy (BLEU) score. The BLEU score evaluates the performance of semantic systems by counting the number of different $m$-grams between sentences, where an $m$-gram is any part of a sentence consisting of $m$ consecutive words. Recently, the most commonly used performance metric has been sentence similarity, which describes the similarity between the original transmitted data and the reconstructed data at the receiver, since it is more suitable for detecting semantic features along whole sentences, in contrast to BLEU. Given a set of $J$ sentences $\mathcal{S} = \{S_1, \cdots, S_j , \cdots, S_J \}$, the sentence similarity is defined as 
\begin{equation}\label{eq:SimDefine}
    M_j = \frac{\mathbf{B}(S_j) \mathbf{B}(S_j')^{T} }{ \|\mathbf{B}(S_j) \| \| \mathbf{B}(S_j')^{T} \|},
\end{equation}
where $(\cdot)^{T}$ denotes the transpose operator of a vector, and $\mathbf{B}(\cdot)$ denotes the bidirectional encoder representations from transformers (BERT) for each sentence $S_j$, which is a vector form representation of the original sentence after its processing by a source-channel encoder and measures the similarity between the transmitted and reconstructed sentence after reception. As observed by \eqref{eq:SimDefine}, some DNN encoding architecture is needed to generate an equivalent vector form of the original data, the most common of which is DeepSC \cite{DeepSC_original}, which aims at reducing the data information to be transmitted while maximizing the mutual information through end-to-end training. 

\begin{figure*}
\centering
\includegraphics[width=0.7\linewidth]{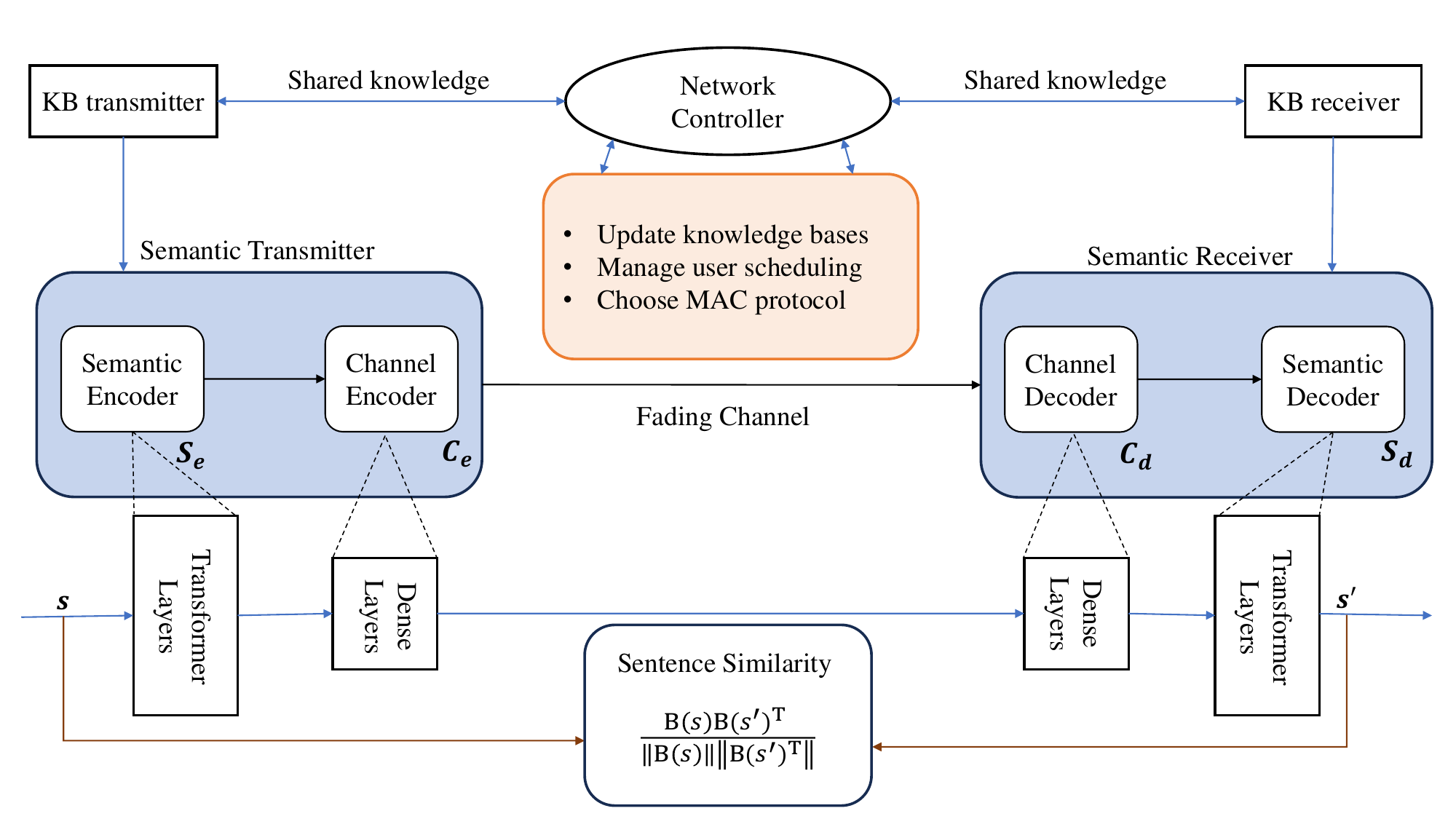}
\caption{Example of semantic network with a DeepSC implemented tranceiver.}
\label{fig:semModel}
\end{figure*}

Given a sentence $S_j$, we denote $S_j'$ its equivalent semantic form, i.e., the output of the semantic encoder part of DeepSC when the input is the given sentence $S_j$. To make the transmitted data resilient to the effects of AWGN and fading, DeepSC encodes $S_j'$ through the channel encoder into a vector $\mathbf{X_j} = [x_{1}, \cdots, x_{kO_{j}}]$, where $\mathbf{X_j}$ consists of the encoded symbols that will be transmitted, $O_{j}$ symbolizes the word number in the $j$-th sentence and $k$ is the number of outputs of the DNN for each word. Since the generated symbols can take arbitrary values, it is clear that DeepSC utilizes discrete-time analog transmission (DTAT), which means that the maximum achievable data rate for it is $C=W$, where $W$ is the allocated bandwidth. Consequently, for $k$ DNN outputs, the number of semantic symbols per sentence is equal to $kO_{j}$ and the corresponding delay when semantic transmission is utilized is given as 
\begin{equation} \label{eq:delaySem}
    D_{l} = \frac{k\sum_{j=1}^{J} {O_{j}}}{{C}}.
\end{equation}
It is straightforward to observe that due to the different number of symbols and capacity of DeepSC, the delay will be different from the one we would get if we had used traditional Shannon communication. From \eqref{eq:delaySem} we can deduce that DeepSC can reduce the delay by appropriately choosing $k$ to meet the desired similarity levels of an application, but cannot achieve the delay reduction of the Shannon paradigm since its rate is fixed.

To make DeepSC more suitable for practical applications, in \cite{DeepSC_lite}, a quantization technique on the constructed unstructured constellation resulting from DeepSC was proposed. A different approach based on knowledge bases and digital transmission instead of end-to-end DNN architectures like DeepSC was also studied in \cite{poor1}, where for energy efficiency reasons, energy harvesting was proposed and the similarity maximization of the system was studied. A variation of sentence similarity, called \textit{semantic accuracy}, has also been studied in \cite{poor2} where energy efficiency is of interest and no energy harvesting is performed.

Regarding multiple access, a NOMA-based system utilizing DeepSC was considered in \cite{seminomaSem} and in \cite{QoEsem} a QoE maximization problem was investigated. Furthermore, in \cite{one_to_many_sem}, the design and use of a DeepSC-like DNN were examined when two users must be served simultaneously. To account for this, the training loss function in this scenario was modified from the original DeepSC to account for both users. Finally, in \cite{Evgenidis_2023} a hybrid system consisting of both digital and semantic transmission was studied as a means to combine the advantages of both techniques. For this purpose an OFDMA technique was considered and delay minimization was studied in combination with sentence similarity constraints. 

\subsection{Semantic communications for image inference}\label{sub:semImage}
Regarding image transmission, the main goal of semantic transmission is the accurate reconstruction of the transmitted image at the receiver side. As in text transmission, the most studied approach for semantic image transmission is through a DNN implementation version of DeepSC, called DeepJSCC, which maps image pixels directly to the channel inputs to be transmitted \cite{deepjscc_gunduz}. Given a sequence of symbols $\mathbf{t}$ representing the original image data and the reconstructed data denoted as $\mathbf{\hat{t}}$, the goal of DeepJSCC is to minimize the distance between the two sequences, which means that the training loss function used is the MSE. To evaluate the performance of image reconstruction, the most appropriate metric is the peak signal-to-noise ratio (SNR), defined as
\begin{equation}\label{eq:psnr}
        \mathrm{PSNR} = 10\log_{10}\left( \frac{\max{\{\mathbf{t}\}}^2 }{||\mathbf{t} - \mathbf{\hat{t}}||^2} \right),
\end{equation}
which gives a measure of the quality of the reconstructed image, with higher peak SNR indicating higher accuracy.

Based on DeepJSCC, there have been several works that aim to improve the performance of the originally proposed model while making it more practical from a usage point of view. In particular, in \cite{deepjscc_q}, a version of DeepJSCC was proposed that quantizes its channel inputs in order to adapt it to existing digital communication systems, and it was shown to have significant performance gains over conventional image transmission techniques. Furthermore, in \cite{paprImage} a peak-to-average power ratio (PAPR) efficient implementation for an OFDM scheme using semantic communications was considered. Apart from these, learned perceptual image patch similarities (LPIPS) have recently been investigated as a means to improve the semantic extraction part of the DNN architecture. In this direction, in \cite{lpips} the perceptual understanding of the reconstructed image data has been studied by optimizing the MSE and LPIPS. In this direction, a collaborative way between conventional and semantic communications was proposed, where the original image data is partially transmitted using conventional source and channel coding techniques and partially using DeepJSCC, while the receiver combines both streams to improve its perceptual similarity between the original and reconstructed data. Similar in concept, but with higher complexity and accuracy, multiple views of the same image are transmitted and a combination of them is used at the receiver end to maximize the peak SNR. 

\subsection{Semantic communications for speech inference}\label{sub:semAudio}
With respect to speech signals, the primary concern of semantic communications is the correct understanding of a speech-transmitted signal by humans. Similar to the text case, a DNN implementation is needed to extract information about the semantics of speech and to perform source-channel coding for robustness against channel fading and noise. As such, an architecture similar to DeepSC, called DeepSC-S, has been proposed for this task \cite{SpSign,SpTran}. If we consider a speech sample sequence $\mathbf{s}$ as the original speech data and $\mathbf{\hat{s}}$ as the output of the speech decoder at the receiver side, the goal of DeepSC-S is to minimize the difference between the original and the reconstructed speech sample sequences. Therefore, DeepSC-C is trained with MSE as the most appropriate loss function. Since the task at hand is the reconstruction of the original speech signals, the performance metric used for such cases is the signal-to-distortion ratio (SDR), which measures the distance between $\mathbf{s}$ and $\mathbf{\hat{s}}$ and is defined as
\begin{equation}\label{eq:SDR}
    \mathrm{SDR} = 10\log_{10}\left( \frac{||\mathbf{s}||^2 }{||\mathbf{s} - \mathbf{\hat{s}}||^2} \right).
\end{equation}
It is easy to observe that the lower the MSE between $\mathbf{s}$ and $\mathbf{\hat{s}}$ is, the larger the SDR will be and consequently better quality at the receiver is achieved. Apart from SDR a useful metric for speech quality recommended by ITU-T is the perceptual evaluation of speech distortion (PESQ), which aims to measure the listening experience of the retrieved speech signals. As such, DeepSC-S has been tested under both performance metrics and  shown to outperform traditional communication schemes for both of them.

A slightly different task that can be of interest when speech signals are involved is speech recognition. In this scenario, text is also of great importance because the received signals are transformed into text data to be compared with the original data. For this purpose, another DNN architecture similar to DeepSC-S was proposed \cite{spRec}, namely DeepSC-SR, which is responsible for transforming the original speech signals into text transcriptions and then performing source-channel coding for communication efficiency. As such, the performance metrics associated with speech recognition are the character error rate (CER) and the word error rate (WER), which measure the difference between the original and retrieved text transcriptions. If we consider a text transcription $\mathbf{t}$, CER and WER are defined as
\begin{equation}\label{eq:CER}
    \mathrm{CER} = \frac{S_C+D_C+I_C}{N_C} \text{ and } \mathrm{WER} = \frac{S_W+D_W+I_W}{N_W}, 
\end{equation}
where $S_{(\cdot)}$, $D_{(\cdot)}$, $I_{(\cdot)}$ represent the number of character/word substations, deletions, insertions, respectively, and $N_{(\cdot)}$ is the number of characters/words in $\mathbf{t}$, respectively.

\subsection{Semantic communications and multiple access}
As discussed, semantic communications can revolutionize the way current communication systems operate by allowing semantic extraction of information and reducing the amount of redundant transmitted data. However, semantic communications have not been extensively studied in multiple access scenarios. Due to the nature of the semantic transceivers proposed so far, i.e., through DNN implementations for encoding, decoding, and information extraction, semantic communications face additional difficulties compared to their conventional counterparts when considering multiple access. In this context, in \cite{one_to_many_sem}, the authors proposed a semantic transceiver design that can support multiple users and also operate for different tasks at the same time, such as text and image transmission. The opposite scenario of a one-to-many broadcast channel was also studied in \cite{one_to_many_sem} and a semantic transceiver was proposed to optimize the performance for all users in the network.

Another challenge that semantic communications must overcome is related to the multimodal nature of the tasks that it can support. Since the semantic transceivers for different tasks each require different DNN architectures and training processes, a device cannot use one transceiver for multiple tasks. With this in mind, in \cite{one_to_many_sem}, a semantic transceiver was designed that can work for different tasks simultaneously. Simulation results showed that such implementations still offer better performance compared to conventional transmission schemes, especially under low SNR, thus highlighting the usefulness of integrating semantic communications in next-generation communication systems.

Despite these major challenges, it is clear that semantic communications are a possible way to provide services for next-generation applications. However, one must consider the practicality of radically changing the existing infrastructure to support them, as well as the fact that they may not always be optimal when traditional KPIs such as delay time are considered apart from the ability of semantic communications to successfully reconstruct the original information. Thus, in \cite{Evgenidis_2023}, a multicarrier system was studied that aims to minimize its transmission delay under strict sentence similarity constraints when both semantic and conventional transmission are possible. As shown, due to the limitations of semantic transceivers to achieve every similarity level, conventional transmission schemes may sometimes be mandatory, while in high SNR values, where Shannon's capacity formula offers a large gain over the fixed data rate of DTAT, conventional schemes are again preferable for transmission. 

\subsection{Future directions}
It is evident that in order to integrate semantic communications into next-generation communication systems, further studies are needed to improve the currently proposed transceiver designs, along with the introduction of multiple access as a fundamental feature in them. To improve the performance of semantic transceivers, it is imperative to consider significant communication KPIs as part of the system. In this sense, a possible direction to explore would be to incorporate application-required KPIs into the training process of designs such as DeepSC to provide joint optimal solutions both for data reconstruction and for meeting application requirements. In this direction, online learning can also be considered as an attractive option to further improve transceiver performance by using the current on-the-fly characteristics of the network to adapt the performance to the new traffic and desired network parameters.

Perhaps one of the most interesting and practical topics for further research is multiple access for semantic communications. One direction that could be exploited is the implementation of special multiple access protocols that support the use of semantic transceivers. The multimodal nature of the different tasks that can be performed could provide another degree of freedom for multiple access. For example, in next-generation systems, the network controller could allow a mix of conventional multiple access protocols, such as TDMA and NOMA for each time slot, where users participating in NOMA could transmit heterogeneous data to take advantage of transceivers that can decode data for different tasks.

\section{ML \& multiple access: A mutual relationship} \label{s:joint}
The field of ML has experienced a significant shift from the dominant ``big data'' strategy, where substantial amounts of data are collected and managed in a central cloud, to a ``small data'' approach \cite{gen2}. In this approach, a collective of agents or devices process their data at the edge of a mobile network. This framework is particularly relevant in scenarios where data is distributed among devices, such as in edge computing configurations or IoT applications. The main incentive for adopting distributed ML is the increasing congestion of wireless channels due to the overwhelming amount of data, which discourages agents from sharing their datasets with ML servers \cite{gen1}.

Consequently, the inherent difficulties associated with wireless networks, such as constrained bandwidth, fluctuating latency, and potential communication breakdowns, pose significant challenges to the smooth integration of ML algorithms. Therefore, in the field of distributed learning, the primary goal of communication shifts from rate maximization to accelerating the training of ML models using distributed data.
The convergence of an ML model now depends on both the selected training algorithm and the utilized multiple access protocol, supporting distributed wireless training of the ML model \cite{gen3}. This facilitates the creation of coherent computation-communication strategies, where the training algorithm is adapted to the communication process of the agents, taking into account factors such as their data volume, the network topology, and the communication pattern of the agents \cite{gen4}. In the following, we present some promising distributed ML frameworks and discuss existing multiple access protocols as well as potential avenues for future research.

\subsection{Federated learning}
Consider the set $\mathcal{N}$ containing all devices that will participate in the FL procedure, with the number of devices equal to $N$. We assume that the $i$-th participating device has its own dataset $\mathcal{D}_i$ of $D_i$ training samples, where each training sample $k$ consists of an input vector $\boldsymbol{x}_{i,k}$ and a target vector $\boldsymbol{y}_{i,k}$. The training goal of the FL procedure is given by 
\begin{equation}\label{eq:fl}
    \underset{\boldsymbol{w}}{\text{min}} \quad \sum_{i\in\mathcal{N}}\frac{d_i}{D_i}\sum_{k\in\mathcal{D}_i}f\left(\boldsymbol{w}, \boldsymbol{x}_{i,k}, \boldsymbol{y}_{i,k} \right),
\end{equation}
where $\boldsymbol{w}$ are the training parameters of the ML model that the devices should jointly construct, $f(\cdot)$ is the loss function that describes the ML task of all devices, and $d_i$ is a scaling parameter with $\sum_{i}d_i=1$. The goal of the FL procedure is to find the vector $\boldsymbol{w}$ that minimizes the loss function of the entire dataset, as given in \eqref{eq:fl} \cite{googleFL}. 

However, the convergence of the FL model cannot be studied independently of the wireless conditions and the capabilities of the devices \cite{FL1, FL2}. Specifically, we define the convergence time of the FL as 
\begin{equation}\label{fl delay}
    T = (T_\mathrm{L}+T_\mathrm{T}) N_\mathrm{T},
\end{equation}
where $T_\mathrm{L}$ is the local ML computation latency at each device, $T_\mathrm{T}$ is the maximum ML model transmission latency, and $N_\mathrm{T}$ is the number of learning steps \cite{FLtime}. Furthermore, the energy consumption of each device is given by 
\begin{equation} \label{flenergy}
    E = (E_\mathrm{L}+E_\mathrm{T}) N_\mathrm{T},
\end{equation}
where $E_\mathrm{L}$ is the energy consumed by each device during a round of local training, and $E_\mathrm{T}$ is the energy consumed during a round of communication \cite{FLenergy}. 

We note that the convergence time of the FL is strongly influenced by the available energy per device and the available bandwidth. It is a common assumption that all devices can upload their local updates during each iteration, while the use of sequential polling with multiple access channels may allow the upload of more local updates at the expense of more bandwidth. However, in practical scenarios, certain users (e.g., mobile phones) may be busy with other tasks or in a sleep state. Consequently, even if they are instructed to upload, they may not be able to access local updates, resulting in channel waste. In addition, from \eqref{fl delay} and \eqref{flenergy}, it is observed that a unified computing and communication design is crucial to reduce FL latency, while also considering FL energy consumption \cite{jointFL, jointFL2, jointFL3, jointFL4, jointFL5}. Thus, multiple access for FL must encompass the balance between loss function convergence, time and energy consumption, and effective management of communication resources \cite{partI}. FL convergence analysis is a critical step in the development of multiple access protocols for FL. The results of FL convergence analysis illustrate the impact of wireless factors on key learning metrics, thus helping to optimize wireless resource allocation and determine other wireless system parameters \cite{FL2, gen2, gen4}. For convergence analysis, it is essential to examine how wireless factors and multiple access protocols affect the convergence of realistic FL with non-convex local ML models and loss functions.
For example, let us consider the result of \cite{QFL}. The authors of \cite{QFL} mention that the upper bound of performance is influenced by the quantization error. 
This indicates that increasing the number of quantization bits reduces the quantization error. However, it was also shown that this approach leads to larger sizes of the local model parameters, potentially causing increased transmission latency. Considering this trade-off between model accuracy and fast convergence, the authors proposed a joint computation and communication design to optimize the OMA protocol used in the FL process and the quantization levels per round. It is noted that wireless fading significantly affects the chosen quantization levels, highlighting the importance of integrated communication and computation design for FL and ML applications.

\subsection{Federated distillation} 
While FL is known for its communication efficiency, it still necessitates the transmission of large models over the air. DNN architectures usually consist of a substantial number of model parameters, making the exchange of such models costly and impeding frequent communication, particularly in environments with limited wireless resources. Conversely, FD only transmits the outputs of the DNN models, which have significantly smaller dimensions compared to the model sizes \cite{FD1}. For instance, in a classification task, each device conducts local iterations while retaining the average model output. These local average outputs are periodically uploaded to the parameter server, which aggregates and averages the local average outputs across devices per class. The resulting global average outputs are then downloaded by each device. To incorporate the downloaded global knowledge into local models, each device undergoes local iterations with its own loss function, in addition to a regularizer that assesses the discrepancy between its prediction output for a training sample and the global average output for the corresponding class of the sample. The communication delay of a device executing FD is detailed in \cite{FD4},
\begin{equation}
    T_\mathrm{L} = \frac{\mathrm{dim}(\boldsymbol{y})Db}{R},
\end{equation}
where $R$ is the transmission rate, $\mathrm{dim}(\boldsymbol{y})$ is the dimension of the DNN's output, $D$ is the batch size, and $b$ is the number of bits used to represent a float number. Compared to FL, which transmits all the parameters of the DNN, the transmission delay caused by transmitting the outputs of the DNN is generally lower, thus the FD has a lower communication overhead \cite{FD1,FD2,FD3,FD4,FD5,FD6,FD7}. 

Nevertheless, in \cite{FD1} it was observed that the performance of FD is slightly worse than that of FL. Thus, there is a trade-off between communication overhead and ML accuracy that is worth investigating. In this context, hybrid FD has been proposed in \cite{FD7}, which combines the concepts of FD and FL to provide a high-accuracy scheme with low communication complexity. However, the convergence analysis of FD has not been widely explored, and most results on FD are based on empirical observations. This limitation can hinder the understanding of the performance of FD under fading conditions and energy consumption constraints. Nevertheless, the joint multiple access and computation design of FD can be studied using the JSCC prism. JSCC does not require a theoretical analysis of FD convergence, yet it can provide an appropriate physical-layer modulation based on fading and the statistics of the information to be transmitted per communication round.

\subsection{Split learning}
SL is another recently emerged collaborative training framework for distributed learning, which is capable of training an ML model between a device and an edge server by splitting the ML model into a device-side model and a server-side model at a cut layer \cite{SL1}. The procedure for SL is as follows: First, the device executes the device-side model with local data and sends intermediate output associated with the cut layer, called smashed data, to the edge server. Subsequently, the edge server executes the server-side model, completing the forward propagation process \cite{SL1, SL2}. Second, the edge server updates the server-side model and sends the gradient of smashed data associated with the cut layer back to the device. The device then updates the device-side model, completing the backward propagation process. This completes the SL process for a device. The updated device-side model is then transferred to the next device, and the process is repeated until all devices are trained. In SL, small-sized device-side models, smashed data, and gradients of smashed data are exchanged between devices and the edge server, resulting in reduced communication overhead compared to uploading the entire AI model in FL \cite{SL3, SL4, SL5}. Due to its superior efficiency, SL is potentially suitable for resource-constrained IoT devices \cite{SL6}. However, when multiple devices participate in SL, all devices interact with the edge server sequentially, resulting in significant training latency, especially when the number of devices is large. The latency of SL is thus strongly influenced by the chosen cut layer and the wireless communication latency \cite{SL9, SL10}.

Split learning (SL) is a recently developed collaborative training approach for distributed learning that facilitates the training of a ML model between a device and an edge server by dividing the ML model into a device-side model and a server-side model at a designated cut layer \cite{SL1}. The SL process unfolds as follows: Initially, the device utilizes the device-side model using local data and transmits the intermediate output related to the cut layer, to the edge server. Subsequently, the edge server uses the server-side model, completing the forward propagation of the DNN model \cite{SL1, SL2}. Next, the edge server updates the server-side model and sends the gradient associated with the cut layer back to the device. The device then updates the device-side model to conclude the backward propagation process, marking the completion of the SL process for a device. The updated device-side model is subsequently passed on to the next device, and this cycle repeats until all devices have been trained. In SL, the exchange of small-sized device-side models, smashed data, and gradients of smashed data between devices and the edge server reduces communication overhead compared to uploading the entire AI model in FL \cite{SL3, SL4, SL5}. Given its efficiency, SL holds promise for resource-constrained IoT devices \cite{SL6}. However, in scenarios where diverse devices wish to participate in SL, all devices communicate with the edge server sequentially, leading to considerable training latency, particularly in cases with a large number of devices. The latency of SL is thus heavily influenced by the selected cut layer and the latency associated with wireless communication \cite{SL9, SL10}.

As a consequence, an interesting trade-off between FL and SL arises, which facilitates the study of joint computation and communication. Let us assume $N$ clients/devices, with an ML model of $K$ parameters, where $D$ is the size of the dataset, $q$ is the size of the smashed layer, $\eta$ is the fraction of model parameters (weights) with the client, and thus $1-\eta$ is the fraction of parameters with the server. Therefore, the communication efficiency, computed as the ratio of the data transfers of FL and SL, is given by \cite{SL1}
\begin{equation} \label{eq:split}
\rho=\frac{2 N D}{2 p q+\eta N D}.
\end{equation}
Thus, SL is better for communication efficiency when $\rho > 1$, while FL is better when $\rho < 1$. By rearranging the terms and expressing them as an equality, and dividing the numerator and denominator by $N D$, we get the equation of a rectangular hyperbola as $N = \frac{2 p q}{(2-\eta) D}$ in the case of client weight sharing, and $N = \frac{p q}{D}$ in the case of no client weight sharing with alternating epochs. This hyperbola separates the areas where one technique performs better than the other. From this example, it is obvious that if both ML methods use the same multiple access protocols, choosing FL or SL based on the equation \eqref{eq:split} can speed up the training latency of the global ML model. 

However, it is interesting to explore whether SL and consequently FL might have different optimal joint computation and communication protocols, given their different training procedures. In such a scenario, \eqref{eq:split} may need to be modified. For example, in \cite{SL6}, an SL scheme called cluster-based parallel SL (CPSL) is designed to reduce training latency by following a ``first-parallel-then-sequential'' approach. CPSL involves partitioning devices into clusters, training device-side models in parallel in each cluster, and then aggregating them. Then, CPSL trains the entire AI model sequentially across clusters, thereby parallelizing the training process and reducing the training latency of SL. A joint computation-communication design is provided by implementing a resource management algorithm to minimize the training latency of CPSL, taking into account device heterogeneity and network dynamics in wireless networks. This is achieved by stochastically optimizing cut layer selection, device clustering, and radio spectrum allocation. While CPSL is shown to outperform an FL benchmark, a detailed comparison between CPSL and an optimally designed FL is still lacking. Furthermore, the interplay between FD and SL has not received much attention. It is also worth noting that the effect of fading and noise on the convergence of SL is unknown. The convergence analysis of SL has not yet been explored, and consequently, a comprehensive analysis of the wireless medium on SL has not been performed. This is critical for real-time IoT applications, where the low-power data transmission of IoT devices is prone to communication errors.

\subsection{Reinforcement learning over wireless networks}

RL allows wireless devices to predict strategies, such as resource management schemes, via trial and error with their underlying wireless environment. RL provides efficient solutions to a wide variety of optimization problems, ranging from Markov decision processes and games to intractable non-convex optimization problems \cite{RL1, RL5, gen3}. Importantly, RL does not require precise modeling of the environment. RL agents can autonomously learn implicit knowledge about network dynamics from raw, high-dimensional observations through interaction with the environment. In multi-agent RL paradigms, agents may need to share various RL information with each other. This could include sharing rewards, RL model parameters, actions, and states among agents. Different collaborative RL algorithms may involve sharing different types of RL information.

Consequently, in wireless networks, the convergence of multi-agent RL depends not only on RL parameters such as the size of the ML model, but also on other factors such as the limited number of resource blocks, the imperfect transmission of RL parameters, and limited transmit power and computational capabilities of devices \cite{RL2, RL3}. In particular, the number of resource blocks determines how many devices can execute the multi-agent RL algorithm. At the same time, fading channels can introduce errors in the RL process. In addition, limited transmit power and computational capabilities significantly affect the time required for RL model updates and RL parameter transmission. Hence, significant hurdles in deploying multi-agent RL across wireless networks are present across several domains. These encompass optimizing resource block allocation and device scheduling for transmitting RL parameters, ensuring reliable and energy-efficient transmission of RL parameters, jointly optimizing RL training methods and wireless resource allocation to minimize RL convergence time, and devising coding and decoding methods.\cite{RL4, RL5}.

\subsection{ML-aided multiple access protocol design}\label{section:mac design}
Recent advances in deep RL (DRL) have generated enthusiasm in the network research community, leading to the exploration of DRL applications in various protocol optimization tasks, including congestion control and multiple access protocol design. The integration of DRL techniques offers the potential to minimize the manual effort required to tune protocol parameters. With numerous ways for a user to communicate with a BS using a given protocol, the application of DRL ensures that wireless networks are not limited to human-designed protocols \cite{mac3}. The upcoming challenge in protocol learning is to enable users and BSs to explore the full range of potential protocols. Communication protocols serve as languages used by network nodes. Therefore, before a user equipment (UE) exchanges data with a BS, it must negotiate the terms and parameters of the transmission. This negotiation involves signaling messages across all layers of the protocol stack. The challenge arises because for two radios to coordinate effectively, they must find a state in which they can interpret each other's messages. Recent research suggests that addressing this common signaling discovery problem in the context of DRL initially involves training the nodes with supervised learning \cite{mac4}. This provides radios with an initial communication protocol, which they can subsequently refine through trial and error.

Essentially, the successful effectiveness of communication between a user and the BS, encompasses various network metrics such as access delay, data rate, and energy efficiency, depends on two key concepts. First, the utilized channel access scheme, which is the method employed by multiple radios to share a communication channel, such as a wireless medium. The channel access scheme is implemented and constrained by the physical layer, with examples including FDMA and TDMA. Second, the multiple access protocol, which encompasses the combination of a channel access policy and signaling used in conjunction with a channel access scheme. Channel access policies, such as listen before talk (LBT) and dynamic scheduling, define how radios access the channel. Signaling serves as the vocabulary and rules, described by the structure of protocol data units, subheaders, etc., that guide radios in coordination. The channel access policy determines when data is sent through the user-plane pipe, while the signaling rules dictate when and what is sent through the control-plane pipe. It is important to note that the end-to-end communication delay of a data transmission is affected by both the signaling overhead and the data rate. Therefore, the joint optimization of control signaling and channel access policies is critical to to enhancing current communication protocols to meet the diverse communication objectives of future networks \cite{mac1, mac2}.

\begin{figure*}
\centering
\includegraphics[width=0.8\linewidth]{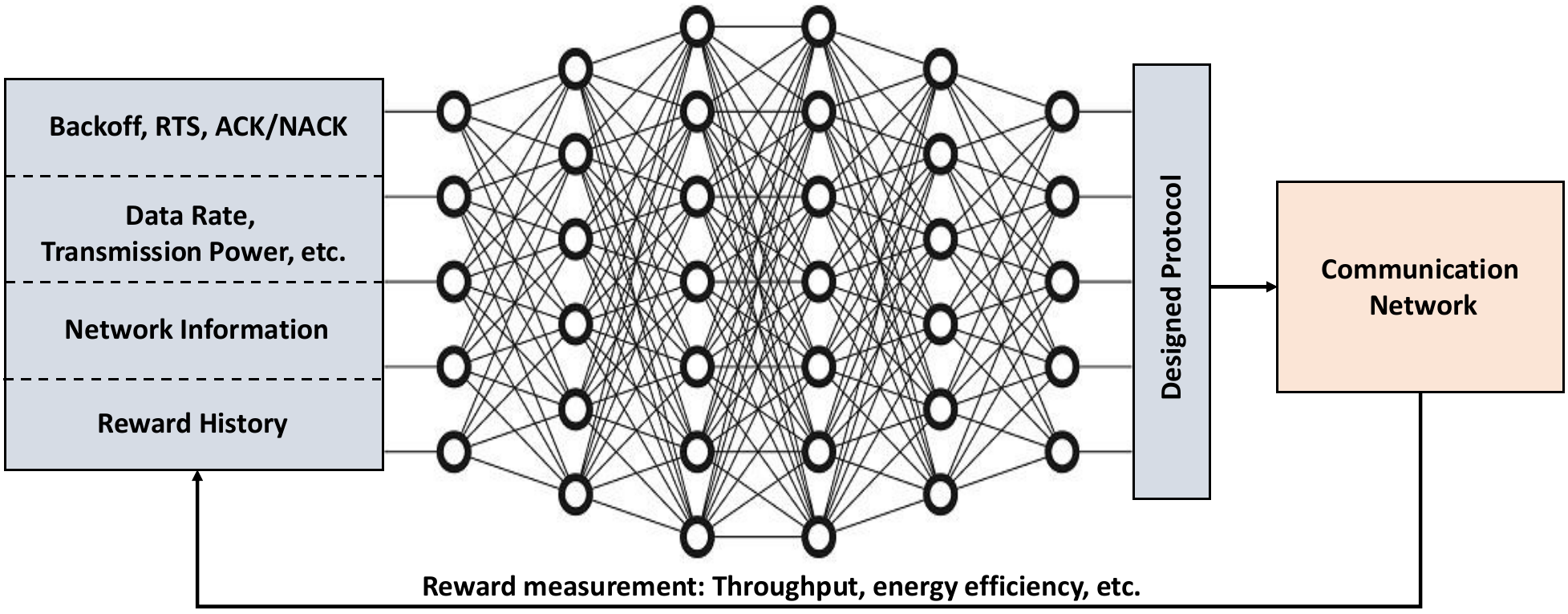}
\caption{RL-aided multiple access protocol design.}
\label{fig:mac}
\end{figure*}

Consider the illustration in Fig. \ref{fig:mac}, where a multiple access protocol design employing deep learning and RL attempts to identify the most effective multiple access protocol from the existing options. Within the DRL framework, it is crucial to define the communication objective that is aligned with specific use cases and device categories across different scenarios. This definition shapes the reward function, a critical component in the optimization process. For instance, in scenarios where IoT devices prioritize maximizing throughput while minimizing energy consumption, the objective function aims for successfully transmitted bits divided by consumed energy per bit. Notably, previous works such as \cite{mac4,mac5,mac6} utilized the average throughput of the link as the reward function, focusing on maximizing the information data rate. In the DRL agent's state-action representation, a vector captures the building blocks, and a history of average link throughput values (for a fixed link) serves as input. The building blocks include various elements such as ACK/NACK usage, back-off mechanisms, listen before talk (LBT), physical layer parameters (e.g., data rate), and the uplink signaling action space (USAS). The USAS, which features larger vocabularies, provides more control information to the channel access policy, providing a richer context to the radio nodes. While this may result in increased signaling overhead, it enables the implementation of more sophisticated channel access algorithms. The size of the signaling vocabulary becomes a crucial hyper-parameter, striking a balance between channel access performance and signaling cost. This delicate equilibrium is exemplified in \cite{mac2}, where the DRL agent adapts to low network load situations by eliminating control packets (e.g., ACK), thereby increasing channel throughput. Conversely, in high-traffic scenarios, the agent opts for protective measures such as ACK, request to sendRTS, clear to send (CTS), smaller frame sizes, and lower bit rates to effectively manage collisions.

Future work could focus on training multiple access protocol learners in networks with a reduced number of UEs and lower traffic volume before deploying them in larger networks. Investigating the mirroring problem of training a BS to learn an expert UE signaling can provide insights into the challenges of protocol learning. The ultimate objective is to jointly train UEs and the BS to evolve an entirely new multiple access protocol. This involves learning not only the signaling and channel access policies, but also the signaling vocabulary. Research in \cite{mac4} suggests that this is challenging when all agents (i.e., UEs and the BS) start with no prior protocol knowledge. Therefore, one potential approach is scheduled training alternating between supervised and unsupervised learning to facilitate the emergence of entirely new protocols. Additionally, exploring multiple access protocol design for random access schemes in mMTC traffic is an intriguing research direction \cite{mac7}.

\subsection{State-of-the-art multiple access protocols}
\subsubsection{OMA} 
OMA protocols, which include techniques such as TDMA and FDMA, play a critical role in distributed ML by facilitating efficient communication between all devices and a central node, if present. In scenarios where devices need to communicate with the central server in a synchronized manner, taking turns to transmit updates, TDMA proves beneficial. TDMA allocates different time slots to each device, preventing interference and ensuring that each device has a dedicated time to transmit its data. On the other hand, FDMA allows devices to transmit their ML model updates on separate frequency bands. This approach prevents interference and allows simultaneous communication. This is especially useful when devices have different available frequency bands, and it can also parallelize the ML training process. In addition, OMA protocols have fewer synchronization issues compared to interference-based protocols such as OTA computing. This is crucial in distributed ML to ensure that all updates from the devices are received and processed in a coordinated manner. For these reasons, the optimization of communication for distributed ML, utilizing OMA transmissions, has been extensively explored in recent literature \cite{floma1,floma2,floma3,floma4,floma5,floma6,FLenergy,FLtime,QFL,FD1,FD2,FD3,FD4,FD5,FD6,FD7}.
\subsubsection{Interference-based protocols}
Computional capabilities, data sizes, and transmission requirements often vary among different devices. Multiple access protocols designed for interference mitigation enable non-orthogonal resource sharing, allowing multiple devices with different requirements and channel conditions to communicate simultaneously \cite{flnoma1, floma4}. In this context, NOMA with its non-orthogonal channel access can improve spectral efficiency in distributed ML scenarios \cite{flnoma2, flnoma3}. Devices can share the same time and frequency resources, allowing more efficient communication of ML models, which is particularly relevant when dealing with a large number of devices. Compared to OMA protocols, NOMA can reduce communication delays, thereby accelerating ML training \cite{flnoma5, flnoma6, flnoma7}. In addition to NOMA, RSMA has found application in distributed ML \cite{flrsma1, flrsma2, flrsma3, flrsma4}. The uplink communication between devices and the central node typically serves as the latency bottleneck for these ML schemes. RSMA involves splitting each agent's data rate into two parts: a common part shared by all users and a private part exclusive to each user. This allows multiple devices to simultaneously transmit their private information while sharing a common message, contributing to increased spectral efficiency. As a result, RSMA has been shown to reduce the training time of distributed ML models compared to NOMA and OMA \cite{flrsma1, flrsma2, flrsma3, flrsma4}. RSMA is also easier to implement in the uplink and is compatible with multiple antennas. Furthermore, the concept of OTA computing has been used to reduce communication delays during the training of distributed ML models, which significantly improves the spectral efficiency of communication during ML training. 

\subsection{Future directions}
\subsubsection{Contention-based protocols}
Contention-based protocols are particularly beneficial in scenarios where devices can dynamically join or leave the distributed ML process, as they help manage communication efficiently with low communication overhead. In addition, these protocols address partial participation by allowing devices to transmit their updates to the central server whenever they are ready, without requiring prior coordination. This requires that the ML training algorithm take partial participation into account during training, and convergence analysis is essential to assess the impact of random access and collisions on the training process. Furthermore, contention-based protocols are valuable for modeling asynchronous communication, where devices transmit their updates independently. This is practical in distributed ML scenarios, as devices may have different computational speeds, data sizes, and transmission capabilities. These protocols are also effective in conserving communication resources in sparse networks, where the number of active devices transmitting at any given time is small. For example, random access can be used in scenarios involving low-power IoT devices at the edge, where network sparsity can be observed. Indeed, in \cite{cb1, cb2, cb3, cb4}, random access was shown to efficiently handle such conditions.

The convergence analysis of FL, FD, SL or distributed RL under random access has not been sufficiently studied, despite the use of state-of-the-art random access protocols. There is an opportunity to propose new random access protocols by appropriately designing their back-off mechanisms, taking into account the computational capabilities of the devices. In addition, incorporating priority mechanisms to manage devices with different priorities in transmitting their updates is an interesting avenue. These priorities can be either delay-based or information-based, assessing the importance of the update from a particular device for the convergence speed of the ML scheme. Furthermore, interference mitigation protocols such as CDMA can be used to reduce the number of collisions without increasing the communication overhead. However, it is important to note that contention-based protocols, including random access, do not guarantee communication delay. This aspect should be carefully considered, especially when investigating these protocols for delay-intolerant ML applications. It is worth noting that random-access protocols as enablers have not been extensively studied in the context of SL, FD, or distributed RL, while they have received minimal attention in the context of FL.

\subsubsection{Adaptive modulation}
While there has been considerable research on improving the communication efficiency and learning performance of wireless FL, the impact of modulation selection on learning performance remains largely unexplored, with exceptions found in \cite{mod1, mod2}. In addition to channel conditions, local processing power and data importance also influence modulation selection. Furthermore, modulation selection has an impact on both learning latency and convergence rate due to transmission rate and symbol error rate (SER). High-order modulation schemes can yield a large SER and a high transmission rate, potentially reducing learning latency at the expense of a slower convergence rate \cite{mod1, mod2}. Conversely, a small modulation order may improve the convergence rate but increase the learning latency. Despite these considerations, adaptive modulation has not received significant attention for ML applications. The joint design of computation and adaptive modulation for future ML applications is critical to the integration of distributed ML applications in future wireless networks.

\subsubsection{Multiple access protocols under communication errors} 
The majority of works on distributed ML typically assume a perfect communication channel between the devices and the central node. In these scenarios, the communication data rate is often modeled using the Shannon capacity formula. However, the real-world implications of unreliable communication on learning performance cannot be underestimated, as SER impacts both communication and convergence outcomes \cite{poor}. It is imperative to investigate strategies for enhancing the learning performance of distributed ML systems operating within unreliable channels. For instance, \cite{pase} examined the impact of packet error rate on FL performance, leading to the formulation of an optimization problem for user scheduling and resource allocation. In a more practical context involving imperfect CSI, the Lyapunov optimization framework was utilized to address joint scheduling and resource allocation concerns in \cite{wadu}. Additionally, \cite{chen} introduced a scheduling scheme tailored for imperfect CSI with the objective of achieving a specific training accuracy, albeit at the expense of devices unable to maintain a predefined transmission rate to expedite convergence time. While these studies offer valuable insights, further research is warranted to explore the ramifications of fading and imperfect knowledge of the communication environment, particularly concerning FD, SL, and multi-agent RL.

\section{Digital twinning} \label{s:DT}
In the context of next-generation multiple access, digital twins play a critical role. Multiple access schemes are essential in communications to allow many users or devices to share the same communication channel. As networks become more complex, especially in the beyond-5G era, managing and optimizing these networks becomes more challenging.
Digital twinning is valuable in this area because it can simulate complex network scenarios. This capability is important for testing and improving multiple access strategies in a controlled, virtual environment. It helps to understand how different devices and users interact within a network and to predict the impact of changes in one part of the system on the entire network. By mirroring the network in a virtual space, digital twinning gives operators insight into where the network is underperforming, where it might fail and where there are opportunities to make it more efficient.
Digital twinning also supports advanced network management concepts such as network slicing and dynamic resource allocation in multi-access networks. With a real-time digital version of the physical network, operators can more effectively allocate resources based on current and predicted demand. This approach increases network efficiency and reliability.

Besides their utilization in applications related to communications, digital twins are considered effective in other use cases. For example, they can be applied in smart factory and industry 4.0, where the goal is to reduce production costs, increase efficiency, and provide companies with an increasingly flexible approach to production. Another use case is smart infrastructure and smart cities, where infrastructure management through digital twinning aims to minimize construction and maintenance costs while ensuring safety and lifetime \cite{Mihai2022}.

\subsection{Proactive resource allocation}
In the context of next-generation multiple access, digital twinning revolutionizes resource allocation by effectively allocating network resources based on current and predicted network demand. Digital twins help make this allocation both efficient and proactive with their real-time data analysis and predictive modeling capabilities.

Digital twins incorporate various parameters such as user density, traffic patterns, and resource utilization to create a dynamic model of the network that accurately reflects its current state. This model is constantly updated with real-time data to provide an up-to-date representation of the state of the network. However, digital twins not only reflect the current state of the network, but also use predictive analysis to forecast future network conditions. By analyzing trends and patterns in data usage, device connectivity and service demand, network operators can anticipate where resources will be needed most before demand actually occurs.

This foresight enables automated adjustments to resource allocation. For example, to prevent congestion or service degradation, the digital twin can proactively reroute resources to a particular network zone if it predicts high demand in that area. Conversely, during periods of low demand, resources can be reduced to save energy and lower operating costs. This level of automation in resource management is critical to maintaining network efficiency and reliability.

In addition, digital twins allow operators to simulate different scenarios and plan resource allocation strategies accordingly. They can model the impact of various events, such as a large public gathering or the introduction of a new high-bandwidth service, and create robust resource allocation plans that can efficiently handle such scenarios. This capability is critical to ensuring that the network is prepared for a wide range of potential situations.

The primary objective of proactive resource allocation through digital twinning is to improve the user experience. By ensuring that network resources are optimally distributed and available exactly where and when they are needed, digital twins help maintain high quality of service, reduce latency and prevent outages. They also provide a feedback loop that continuously monitors the results of resource allocation decisions, allowing the network to adapt to changing demands and usage patterns.
To this end, digital twinning can be utilized to provide proactive resource allocation in various scenarios, such as open RAN (O-RAN) \cite{Masaracchia2023}, smart grids \cite{Othman2023}, wireless power transfer \cite{Lv2022}, among others.

\subsection{Digital twinning in MEC}
Many works on digital twinning for wireless communication focus on designing digital twins for MEC servers \cite{Zhang2022,Dai2021,Lu2021,Dong2019,Sun2020,DoDuy2022,Zhou2021}, where the MEC digital twin is used to extract optimal network orchestration in real time. For example, in \cite{Dai2021,Lu2021,Dong2019}, algorithms based on DRL were investigated for training the digital twin of the MEC network to make offload decisions, edge and resource allocation, increase energy efficiency, and reduce migration costs. Moreover, \cite{Sun2020,DoDuy2022} used the digital twin of a MEC system to minimize the end-to-end latency by also taking into account the deviations between the two twins. Furthermore, digital twinning has also been proposed for use in the internet of vehicles \cite{Hu2022,Sun2022} to enable efficient resource allocation policies of vehicles and real-time traffic prediction. However, an important issue is that the digital twin may not always accurately reflect that of its real-world counterpart, since the latter may be subject to rapid changes along with the evolution of time. Therefore, it is necessary to synchronize the digital twin with its physical counterpart to ensure that its state is accurate. To address this challenge, a dynamic hierarchical framework has been designed in \cite{Han2022}, where a cluster of IoT devices is used to sense and collect status information of physical objects to support the update of the digital twin.

\subsection{Multiple access for digital twinning}
Selecting an appropriate access scheme is critical when designing the wireless interface for digital twinning to enable communication between the digital twin and physical entities. In this direction, several schemes can be used to achieve wireless access, such as OFDMA, TDMA, CDMA, and NOMA \cite{Khan2022}.

OFDMA, which uses orthogonal resource blocks for communication, is characterized by low complexity in receiver design. However, it tends to have low spectral efficiency and can only accommodate a limited number of users. NOMA, on the other hand, allows all users to use the entire available bandwidth simultaneously, with decoding at the receiver based on received power levels. This method has high spectral efficiency, supports high connection density, and improves user fairness. However, the trade-off is increased complexity at the receiver and sensitivity to channel uncertainty.

CDMA is known for providing better coverage and operating efficiently at low transmit power levels. Despite these advantages, it faces challenges, especially in the asynchronous wideband CDMA variant, which is associated with high complexity. In addition, channel equalization in CDMA tends to be more complex than in TDMA, and there is a need for guard time between adjacent channels to prevent interference.

On the other hand, TDMA generally has less complex channel equalization requirements than CDMA and does not require a frequency guard band. However, it does require guard bands between adjacent channels to minimize crosstalk and has a high synchronization overhead, which can be a drawback in certain network configurations.

To this end, it should be noted that energy efficiency is also a critical factor in the air interface design for digital twinning. The choice of access scheme has a significant impact on the energy consumption of the network, which affects both the cost of ownership and the environmental footprint of the system.

\section{Conclusion} \label{s:concl}
In this work, the role of NGMA in communication and computation has been explored, focusing on two main objectives: IRCE and IRDCE. Although both objectives aim to use communication and computing resources more effectively, they measure success differently. IRCE focuses on meeting communication quality requirements, including factors such as efficiency, speed, reliability and connectivity. IRDCE, on the other hand, looks at achieving computing goals, such as processing more data either locally or at the edge of the network, or reducing errors in data analysis.

We first examined MEC, which is key to meeting the growing need for data processing and computing power close to the user, and also discussed network slicing. We then explored OTA computing, which is promising for the rapid processing of various functions, particularly useful for immediate computing needs. We also discussed semantic communications, which aims to make communication systems more efficient by focusing on sending meaningful information and reducing unnecessary data. We also looked at how ML interacts with multiple access technologies, highlighting the importance of FL, FD, SL, RL, and ML-based protocol design for multiple access. Finally, we discussed digital twinning and its use in network management, showing how creating virtual models of physical networks can help make networks more efficient and reliable.

\bibliographystyle{IEEEtran}
\bibliography{bib}

\begin{IEEEbiography}[{\includegraphics[width=1in,height=1.25in,clip,keepaspectratio]{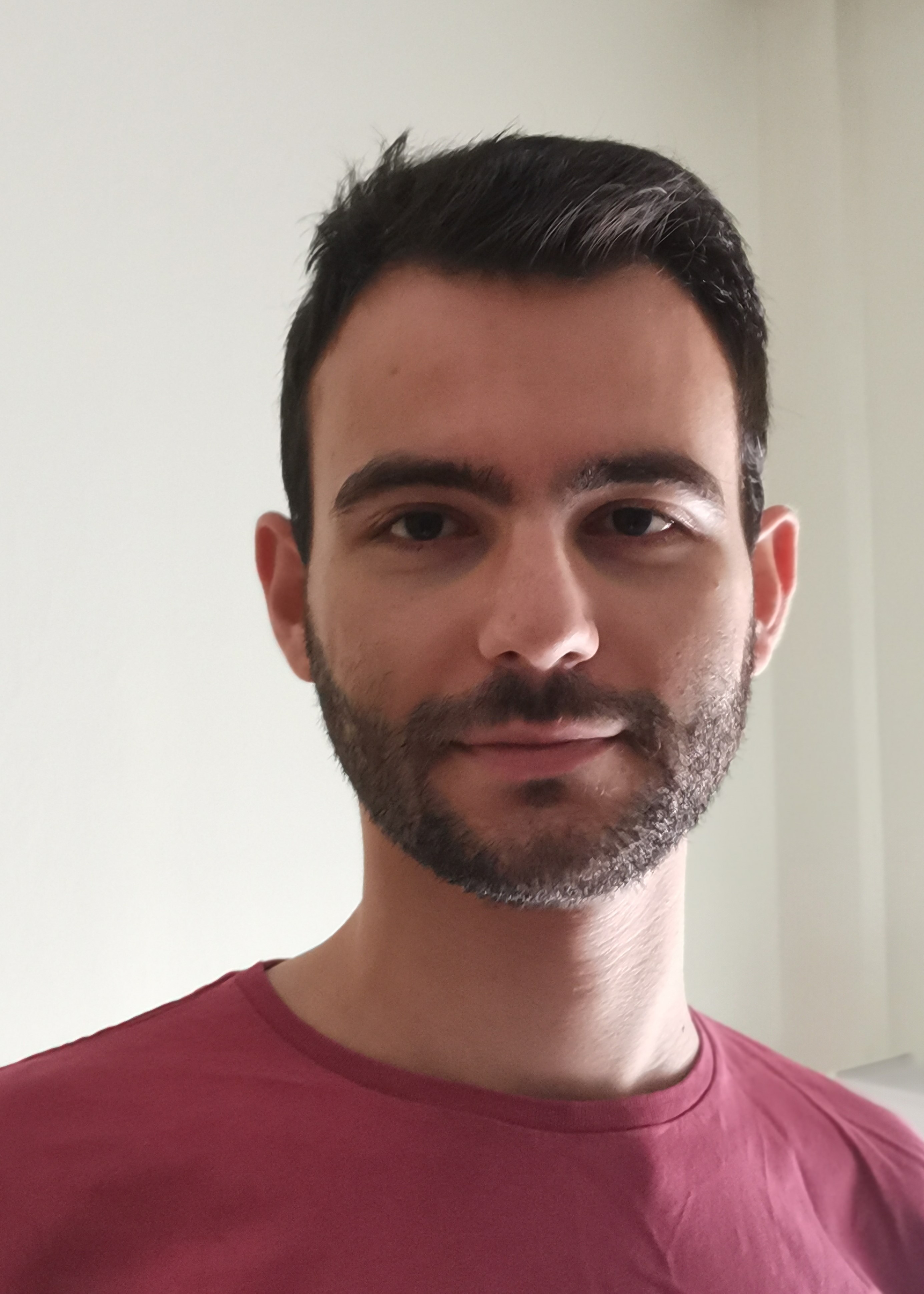}}] {Nikos G. Evgenidis}  received the Diploma (5 years) in Electrical and Computer Engineering from the Aristotle University of Thessaloniki, Greece, in 2022, where he is currently pursuing his PhD. He is also a member of the Wireless and Communications and Information Processing (WCIP) group. His major research interests include semantic communications, over-the-air computing, non-orthogonal multiple access, machine learning and optimization theory.
\end{IEEEbiography}

\begin{IEEEbiography}
[{\includegraphics[width=1in,height=1.25in,clip,keepaspectratio]{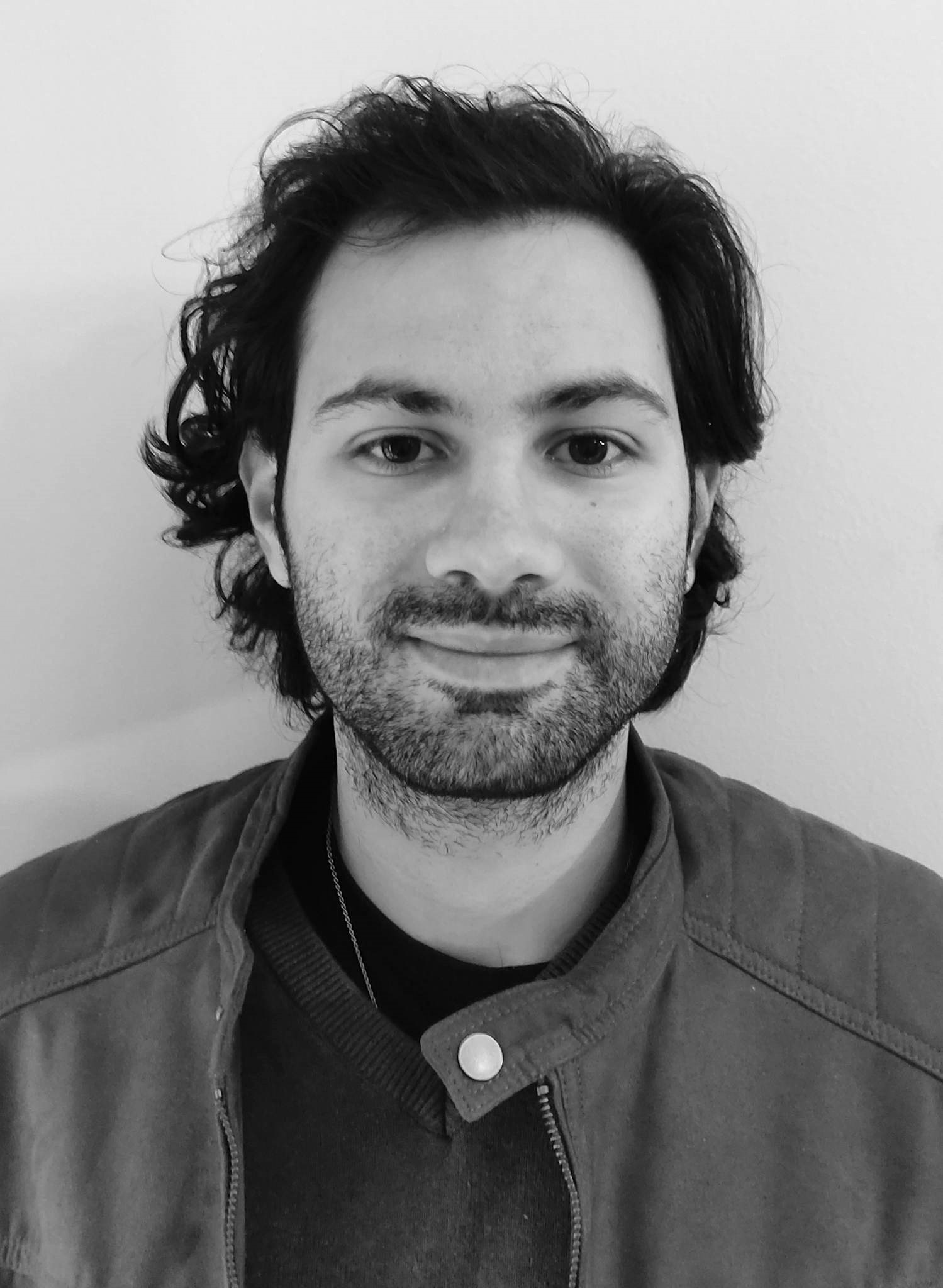}}]{Nikos A. Mitsiou}~(Graduate Student Member, IEEE) received the Diploma Degree (5 years) in Electrical and Computer Engineering from the Aristotle University of Thessaloniki (AUTH), Greece, in 2021, where he is currently pursuing his PhD with the Department of Electrical and Computer Engineering. He is a member of the Wireless and Communications \& Information Processing (WCIP) Group. He was an Exemplary Reviewer of the IEEE Wireless Communications Letters in 2022 (top 3\% of reviewers). His research interests include optimization theory, and machine learning with application to wireless networks.
\end{IEEEbiography}

\begin{IEEEbiography}[{\includegraphics[width=1in,height=1.25in,clip,keepaspectratio]{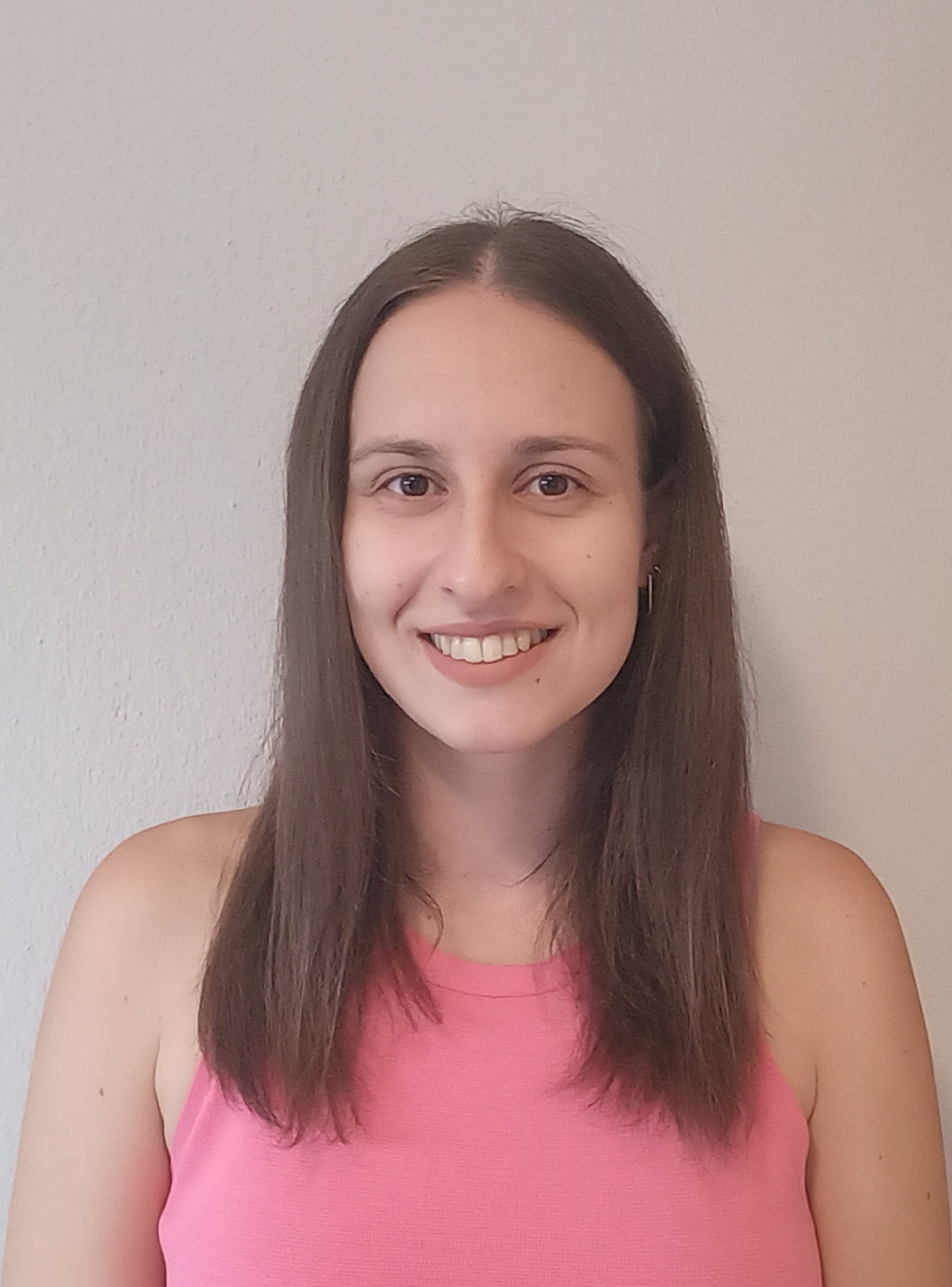}}] {Vasiliki I. Koutsioumpa} received the Diploma (5 years) in Electrical and Computer Engineering from the Aristotle University of Thessaloniki, Greece, in 2023, where she is currently pursuing her PhD. She is also a member of the Wireless Communications and Information Processing (WCIP) Group. Her major research interests include next-generation multiple access schemes, satellite networks and optimization theory.
\end{IEEEbiography}

\begin{IEEEbiography}
[{\includegraphics[width=1in,height=1.25in,clip,keepaspectratio]{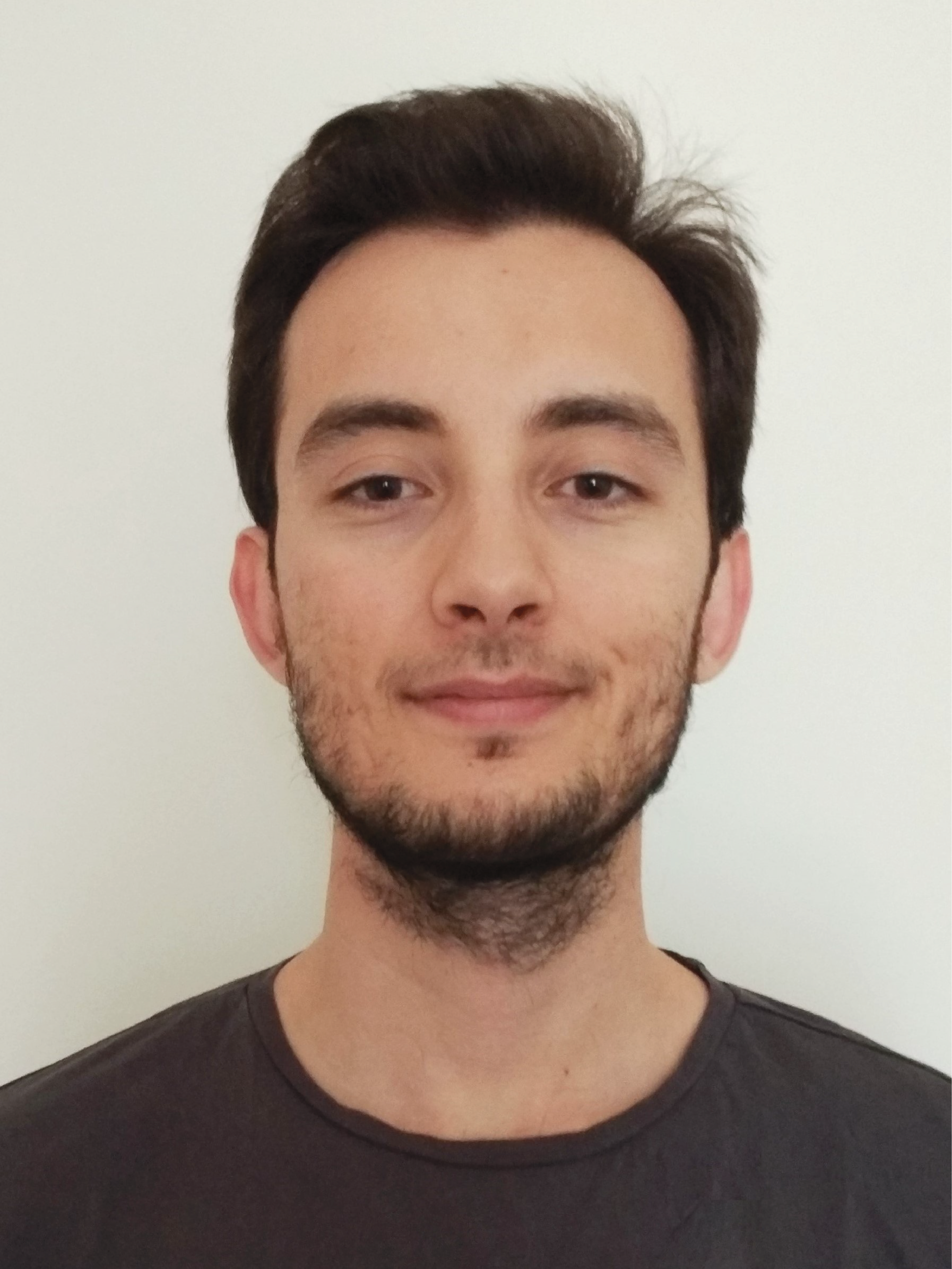}}]{Sotiris A. Tegos } (Senior Member, IEEE) received the Diploma (5 years) and Ph.D. degrees from the Department of Electrical and Computer Engineering, Aristotle University of Thessaloniki, Thessaloniki, Greece, in 2017 and 2022, respectively. Since 2022, he is a Postdoctoral Fellow at the Wireless Communications and Information Processing (WCIP) Group, Aristotle University of Thessaloniki, Thessaloniki, Greece, and at the Department of Applied Informatics, University of Macedonia, Thessaloniki, Greece. Since 2023, he is also a Postdoctoral Fellow at the Department of Electrical and Computer Engineering, University of Western Macedonia, Kozani, Greece. In 2018, he was a visiting researcher at the Department of Electrical and Computer Engineering, Khalifa University, Abu Dhabi, UAE. His current research interests include multiple access in wireless communications, optical wireless communications, and reconfigurable intelligent surfaces. He is a Working Group Member of the Newfocus COST Action “European Network on Future Generation Optical Wireless Communication Technologies”. He serves as an Editor for IEEE Communications Letters. He received the Best Paper Award in 2023 Photonics Global Conference (PGC). He was an exemplary reviewer in IEEE Wireless Communications Letters in 2019 and 2022 (top 3\% of reviewers).
\end{IEEEbiography}
\begin{IEEEbiography}[{\includegraphics[width=1in,height=1.25in,clip,keepaspectratio]{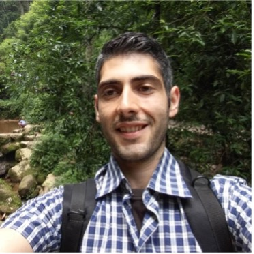}}]{Panagiotis D. Diamantoulakis } (Senior Member, IEEE) received the Diploma (5 years) and the Ph.D. degree from the Department of Electrical and Computer Engineering, Aristotle University of Thessaloniki, Thessaloniki, Greece, in 2012 and 2017, respectively. Since 2022 he is a Postdoctoral Fellow with the Department of Applied Informatics, University of Macedonia, Thessaloniki, Greece. Since 2017, he has been a Postdoctoral Fellow with Wireless Communications and Information Processing (WCIP) Group, AUTH and since 2021, he has been a Visiting Assistant Professor with the Key Lab of Information Coding and Transmission, Southwest Jiaotong University, Chengdu, China. His research interests include optimization theory and applications in wireless networks, optical wireless communications, and goal-oriented communications. He is a Working Group Member of the Newfocus COST Action “European Network on Future Generation Optical Wireless Communication Technologies.” He serves as an Editor of IEEE Open Journal of the Communications Society, Physical Communications (Elsevier), and Frontiers in Communications and Networks, while during 2018-2023 he has been an Editor of IEEE Wireless Communications Letters. He was also an Exemplary Editor of the IEEE Wireless Communications Letters in 2020, and an Exemplary Reviewer of the IEEE Communications Letters in 2014 and the IEEE Transactions on Communications in 2017 and 2019 (top 3\% of reviewers).
\end{IEEEbiography}
\begin{IEEEbiography}[{\includegraphics[width=1in,height=1.25in,clip,keepaspectratio]{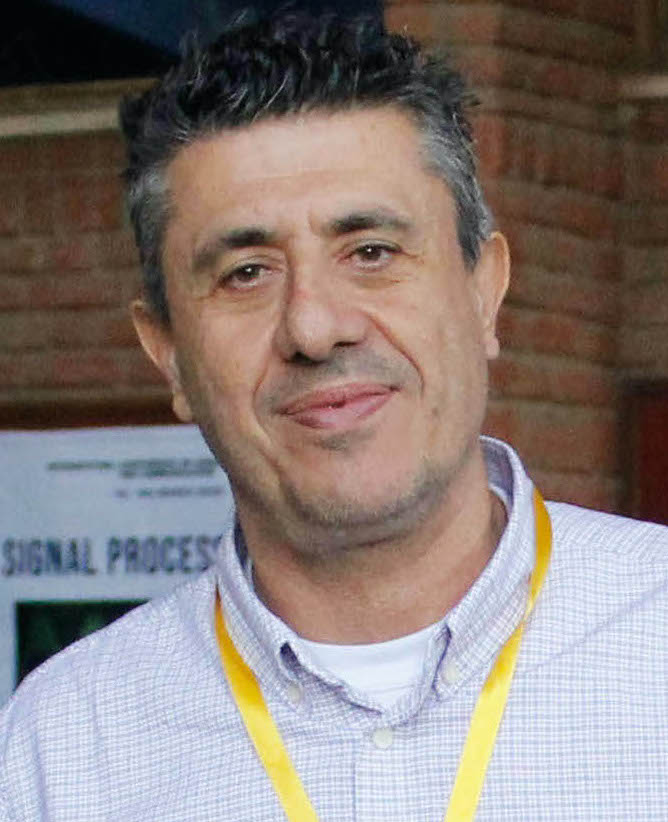}}]{George K. Karagiannidis } (Fellow, IEEE) is currently Professor in the Electrical \& Computer Engineering Dept. of Aristotle University of Thessaloniki, Greece and Head of Wireless Communications \& Information Processing (WCIP) Group. He is also Faculty Fellow in the Cyber Security Systems and Applied AI Research Center, Lebanese American University. His research interests are in the areas of Wireless Communications Systems and Networks, Signal processing, Optical Wireless Communications, Wireless Power Transfer and Applications and Communications \& Signal Processing for Biomedical Engineering.
Dr. Karagiannidis was in the past Editor in several IEEE journals and from 2012 to 2015 he was the Editor-in Chief of IEEE Communications Letters. From September 2018 to June 2022 he served as Associate Editor-in Chief of IEEE Open Journal of Communications Society. Currently, he is the Editor-in-Chief of IEEE Transactions on Communications.
Recently, he received three prestigious awards: The 2021 IEEE ComSoc RCC Technical Recognition Award, the 2018 IEEE ComSoc SPCE Technical Recognition Award and the 2022 Humboldt Research Award from Alexander von Humboldt Foundation.
Dr. Karagiannidis is one of the highly-cited authors across all areas of Electrical Engineering, recognized from Clarivate Analytics as Highly-Cited Researcher in the nine consecutive years 2015-2023.
\end{IEEEbiography}
\end{document}